\documentstyle[epsfig,twoside,12pt]{article}

\catcode`\@=11
\renewcommand{\ps@myheadings}{\let\@mkboth\@gobbletwo
\def\@oddhead{\hbox{}\hfil {\normalsize\sl\rightmark}\hfil 
{\bf\arabic{page}}\hbox{}}
\def\@oddfoot{}\def\@evenhead{{\bf\arabic{page}}\hfil  
{\normalsize\sl\leftmark} \hfil\hbox                  
{}}\def\@evenfoot{}\def\sectionmark##1{}\def\subsectionmark##1{}}

\newcommand{\yes}{yes}
     
\newcommand{\labelshow}{no}        

\newcommand{\labelformat}[1]              
           {\{#1\}}     

\newcommand{\tag}[1]{%
\ifx\labelshow\yes
\quad\mbox{\labelformat{#1}}\protect\label{eq:#1}%
\else%
\protect\label{eq:#1}
\fi}

\newcommand{\eq}[1]{%
\ifx\labelshow\yes
Eq.\ {\{#1\}}(\ref{eq:#1})%
\else%
Eq.\ (\ref{eq:#1})
\fi}

\newcommand{\eqnoeq}[1]{%
\ifx\labelshow\yes
{\{#1\}}(\ref{eq:#1})%
\else%
(\ref{eq:#1})
\fi}

\newcommand{\rangeref}[2]{%
\ifx\labelshow\yes
Eqns.\ ({\{#1\}}%
\ref{eq:#1})--({\{#2\}}%
\ref{eq:#2})
\else%
Eqns.\ (\ref{eq:#1})--(\ref{eq:#2})\nobreak   
\fi}

\newcommand{\reference}[1]{%
\ifx\labelshow\yes
{\{#1\}}\ \cite{#1}%
\else%
\cite{#1}
\fi}

\newcommand{\figgy}[4]{%
\ifx\labelshow\yes
\begin{figure}[tbp]
\vspace{#2}%
\begin{center}\parbox{5 true in}{%
\caption[]{\{#1\} \protect\label{fg:#1} #4}%
}\end{center}%
\end{figure}%
\else
\begin{figure}[tbp]
\vspace{#2}%
\begin{center}\parbox{5 true in}{%
\caption[]{\small \protect\label{fg:#1} #4}%
}\end{center}%
\end{figure}%
\fi}

\newcommand{\twofiggy}[8]{%
\ifx\labelshow\yes
\begin{figure}[tbp]
\vspace{#2}%
\begin{center}\parbox{5 true in}{%
\caption[]{\small \{#1\} \protect\label{fg:#1} #4}%
}\end{center}%
\vspace{#6}%
\begin{center}\parbox{5 true in}{%
\caption[]{ \small \{#5\} \protect\label{fg:#5} #8}%
}\end{center}%
\end{figure}%
\else
\begin{figure}[tbp]
\vspace{#2}%
\begin{center}\parbox{5 true in}{%
\caption[]{ \small \protect\label{fg:#1} #4}%
}\end{center}%
\vspace{#6}%
\begin{center}\parbox{5 true in}{%
\caption[]{\small \protect\label{fg:#5} #8}%
}\end{center}%
\end{figure}%
\fi}

\def\picture #1 by #2 (#3){ 
  \vbox to #2{
    \hrule width #1 height 0pt depth 0pt
    \vfill
  \special{picture #3} } }

\def\centerpicture #1 by #2 (#3 scaled #4){
  \dimen0=#1 \dimen1=#2 
  \divide\dimen0 by 1000 \multiply\dimen0 by #4
  \divide\dimen1 by 1000 \multiply\dimen1 by #4
	\ashift{\dimen0}
	\picture \dimen0 by \dimen1 (#3 scaled #4)}

\newlength{\alen}
\newcommand{\ashift}[1]{
\setlength{\alen}{\textwidth}
\addtolength{\alen}{-1#1}
\addtolength{\alen}{-0.5\alen}
\addtolength{\alen}{-8 pt} \hspace{\alen}}

\newcommand{\alignc}{c}
\newcommand{\isok}{c}

\newcommand{\fgcaption}[3]{\renewcommand{\isok}{#1}%
 \ifx  \alignc \isok
\begin{center}%
\parbox{4 in}{\begin{center}%
{\small{\bf\noindent Fig.\  \ref{fg:#2}\hspace{1em}} #3}  \end{center}}%
\end{center}%
\else%
\begin{center}%
\parbox{4 in}{%
{\small{\bf \noindent Fig.\  \ref{fg:#2}\hspace{0.5em}} #3}}%
\end{center}%
\fi}

\newcommand{\fig}[1]{%
\ifx\labelshow\yes
Fig.~{ \{#1\}} \ref{fg:#1}%
\else%
Fig.~\ref{fg:#1}
\fi}

\newcommand{\figs}[2]{%
\ifx\labelshow\yes
Figs.~{ \{#1\}} \ref{fg:#1}--{ \{#2\}} \ref{fg:#2}%
\else%
Figs.~\ref{fg:#1}--\ref{fg:#2}
\fi}

\newcommand{\tableinsert}[5]{
\ifx\labelshow\yes
\begin{table}
\begin{center}\parbox{5 true in}{%
\caption[]{\{#1\} \protect\label{tb:#1}#4}%
}\end{center}%
\par\vspace {#2}%
\noindent{\footnotesize #5}%
\end{table}%
\else%
\begin{table}
\begin{center}\parbox{5 true in}{%
\caption[]{\protect\label{tb:#1}#4}%
}\end{center}%
\par\vspace {#2}%
\noindent{\small #5}%
\end{table}%
\fi}

\newcommand{\tableref}[1]{%
\ifx\labelshow\yes
Table~{\{#1\}} \ref{tb:#1}%
\else%
Table~\ref{tb:#1}
\fi}%

\newcommand{\heading}[2]{%
\ifx\labelshow\yes
\section{\{#1\} \protect\label{h:#1}#2}%
\else%
\section{\protect\label{h:#1}#2}%
\fi}%

\newcommand{\subheading}[2]{%
\ifx\labelshow\yes
\subsection{\{#1\} \protect\label{sh:#1}#2}%
\else%
\subsection{\protect\label{sh:#1}#2}%
\fi}%

\newcommand{\subsubheading}[2]{%
\ifx\labelshow\yes%
\subsubsection{\{#1\} \protect\label{ss:#1}#2}
\else%
\subsubsection{\protect\label{ss:#1}#2}
\fi}

\newcommand{\align}[1]
{\hskip-0.5em&#1&\hskip-0.5em}          
\newcommand{\flush}{&&\hspace{-1ex}}    

\newcommand{\tightdot}{\kern-.2em\cdot\kern-.2em}

\newcommand{\halfthin}{\kern 0.0834em}
\newcommand{\neghalfthin}{\kern -0.0834em}

\newcommand{\quarterthin}{\kern 0.0417em}
\newcommand{\negquarterthin}{\kern - 0.0417em}

\newcommand{\pdagger}{\vphantom{^\dagger}}

\newcommand{\isotope}[2]{$^{#1}$#2}

\newcommand{\su}[1]{\mbox{$SU_{#1}$}}
\def\Sp#1{\mbox{$S\negquarterthin p_{\quarterthin#1}$}}
\newcommand{\so}[1]{\mbox{$SO_{#1}$}}

\newcommand{\tsub}[1]{_{\mbox{\scriptsize#1}}}
\newcommand{\tsup}[1]{^{\mbox{\scriptsize#1}}}

\newcommand{\runningheads}[2]{\markboth{\hfill #1\hfill}{\hfill #2\hfill}}

\newcommand{\utk}[1]{\vbox{
     \small\em
     \hfil $^{#1}$Department of Physics, University of Tennessee,
     \hfil \newline\hfil 
     Knoxville, TN 
     37996--1200 USA\hfil}}

\newcommand{\drexel}[1]{\vbox{
     \small\em
     \hfil $^{#1}$Department of Physics and Atmospheric Sciences, 
     \hfil \newline\hfil 
     \small\em Drexel 
     University, Philadelphia, PA 19104 USA
     \hfil}}

\newcommand{\ornl}[1]{\vbox{
     \small\em 
      \hfil $^{#1}$Physics Division, Oak Ridge 
      National Laboratory\hfil \newline \hfil 
      \small\em
      Oak Ridge, TN 37831 USA\hfil}}

\newcommand{\taiwan}[1]{\vbox{
     \small\em
     \hfil $^{#1}$Department of Physics, Chung Yuan Christian University,
     \hfil \newline\hfil 
     Chung-Li, Taiwan 32023 ROC\hfil}}

\newcommand{\utkblurb}{Nuclear physics research at the University of 
                      Tennessee is supported by the U.~S. Department of 
                      Energy through Contracts No.\ DE--FG05--87ER40361 and
                      DE--FG05--93ER40770. }

\newcommand{\ornlblurb}{Oak Ridge National Laboratory is managed by Lockheed
                        Martin
                        Energy Systems, Inc.\ for the U.~S. 
                        Department of Energy under Contract No.\ 
                        DE--AC05--84OR21400. }

\newcommand{\drexelblurb}{Nuclear physics research at Drexel University is
                         supported by the National Science Foundation. }

\newcommand{\taiwanblurb}{Nuclear physics research at Chung Yuan Christian 
                         University is supported by the National Science Council 
                          of the {\small ROC}. }

\newcount\hours
\newcount\minutes
\newcount\tmp
\hours=\time
\divide\hours by 60
\tmp=\hours
\minutes=\time
\multiply\tmp by 60
\advance\minutes by -\tmp
\newcommand{\timenow}{\number\hours:\ifnum\number\minutes>9%
{\number\minutes}\else%
0\number\minutes\fi}

\newcommand{\datefile}{\date{\vspace{10pt}
            \small\sc  
            ---\thinspace Printed from File: {\small
           \lowercase{\jobname}.tex at \timenow}
            \thinspace---\vphantom{\bigg[} 
            \\ 
             \small ---\today---}}

\newcommand{\smalldate}{\date{\vspace{10pt}
             \small ---\today---}}

\newcommand{\mnote}[1]
{\setlength{\marginparwidth}{50pt}%
\setlength{\marginparsep}{10pt}%
\marginpar{\scriptsize\em#1}}

\def\pointsize@{10}
\def\fivepoint{\def\pointsize@{5}%
 \normalbaselineskip7\p@
 \abovedisplayskip7\p@ plus1.8\p@ minus5.4\p@
 \belowdisplayskip7\p@ plus1.8\p@ minus5.4\p@
 \abovedisplayshortskip\z@ plus1.8\p@
 \belowdisplayshortskip2.6\p@ plus1.8\p@ minus1.0\p@
 \textfont@\rm\fiverm
 \textfont@\it\fivei
 \textfont@\bf\fivebf
 \ifsyntax@\def\big##1{{\hbox{$\left##1\right.$}}}\else
 \let\big\eightbig@
 \textfont\z@\fiverm \scriptfont\z@\fiverm \scriptscriptfont\z@\fiverm
 \textfont\@ne\fivei \scriptfont\@ne\fivei \scriptscriptfont\@ne\fivei
 \textfont\tw@\fivesy \scriptfont\tw@\fivesy \scriptscriptfont\tw@\fivesy
 \textfont\thr@@\tenex \scriptfont\thr@@\tenex \scriptscriptfont\thr@@\tenex
 \textfont\itfam\fivei
 \textfont\bffam\fivebf \scriptfont\bffam\fivebf \scriptscriptfont\bffam\fivebf
 \fi
 \setbox\strutbox\hbox{\vrule height 5\p@ depth2\p@ width\z@}%
 \setbox\strutbox@\hbox{\vrule height 4\p@ depth1\p@ width\z@}%
 \normalbaselines\fiverm\ex@=.2326ex}

\def\boxit#1{\vbox{\hrule\hbox{\vrule\kern3pt
     \vbox{\kern3pt#1\kern3pt}\kern3pt\vrule}\hrule}}

\def\yspacef{\kern 0.6\pointsize@\p@}
\def\yspaceo{\kern 0.3em }

\def\cstok#1{
\if5\pointsize@ \let\yspace\yspacef \else
 \let\yspace\yspaceo \fi
  \leavevmode\hbox{\kern-.4pt\vrule\vtop{\vbox{\hrule
      \kern 0.1\pointsize@\p@
        \hbox{\vphantom{\rm/}\yspace{#1}\yspace}}
      \kern 0.1\pointsize@\p@\hrule}\vrule}}

\def\Young#1{\null\,\vcenter{\normalbaselines\m@th
     \if8\pointsize@ \baselineskip=8pt \lineskip=-0.4pt
     \else \baselineskip=10pt \lineskip=-0.4pt \fi
     \if5\pointsize@ \baselineskip=5pt \lineskip=-0.4pt\fi
    \ialign{$##$&&$##$\crcr
      \mathstrut\crcr\noalign{\kern-\baselineskip}
      #1\crcr\mathstrut\crcr\noalign{\kern-\baselineskip}}}\,}

        \newdimen\figheight
        \newdimen\figmove

\newcommand{\texturesepsfiggy}[8]{%
\begin{figure}[t]%
\figheight=#2
\divide\figheight by 1000 \multiply\figheight by #6%
\vspace*{#3}%
\vspace{\figheight}%
\hbox{\hskip #7
\special{illustration #5 scaled #6}%
}%
\begin{center}%
\parbox{5 true in}{%
\caption[\protect{ #8}]{\small \protect\label{fg:#1} #8}%
}
\end{center}%
\vspace{#4}%
\end{figure}%
}%
\newcommand{\orphepsfiggy}[8]{%
\begin{figure}[t]%
\vspace*{#3}
\figheight=#2
\divide\figheight by 1000 \multiply\figheight by #6%
\figmove=#7
\advance\figmove by -1 true in%
\vspace*{#3}%
\hbox{\hskip \figmove
\epsfig{file=#5,height=\figheight}%
}%
\begin{center}%
\parbox{5 true in}{%
\caption[\protect{ #8}]{\small \protect\label{fg:#1} #8}%
\vspace*{#4}}
\end{center}%
\end{figure}%
}%
\newcommand{\orphepsfiggyhere}[6]{%
\vspace*{20pt}
\vspace*{#2}
\figheight=#1
\divide\figheight by 1000 \multiply\figheight by #5%
\figmove=#6
\advance\figmove by -1 true in%
\vspace*{#3}
\hbox{\hskip \figmove
\epsfig{file=#4,height=\figheight}%
}%
}%
\newcommand{\altepsfiggy}[9]{%
\begin{figure}[htbp]%
\figheight=#2
\divide\figheight by 1000 \multiply\figheight by #6%
\figmove=#7
\advance\figmove by -1 true in%
\vspace*{#3}%
\hbox{\hskip \figmove
\epsfig{file=#5,height=\figheight}%
}%
\begin{center}%
\parbox{5 true in}{%
\caption[#9]{\small \protect\label{fg:#1} #8}%
}
\end{center}%
\end{figure}%
}%

\newcommand{\noepsfiggy}[6]{%
\begin{figure}[htbp]%
\figheight=#2
\divide\figheight by 1000 \multiply\figheight by #5%
\vspace*{#3}%
\vspace{\figheight}%
\begin{center}%
\parbox{5 true in}{%
\caption[\protect{ #6}]{\small \protect\label{fg:#1} #6}%
}%
\end{center}%
\vspace{#4}%
\end{figure}%
}%
\newcommand{\orphtex}{orphtex}
\newcommand{\mactex}{mactex}
\newcommand{\nofigtex}{nofigtex}

\newcommand{\wherethehellami}{nofigtex}		

\newcommand{\orph}{%
     \renewcommand{\wherethehellami}{orphtex}}  

\newcommand{\epsfiggy}[8]{%
\ifx\wherethehellami\mactex
  \message{...(epsfiggy) Using LaTeX with TeXtures on a Macintosh.}%
  \texturesepsfiggy{#1}{#2}{#3}{#4}{#5}{#6}{#7}{#8}%
     \else%
\ifx\wherethehellami\orphtex
  \message{...(epsfiggy) Using LaTeX on orph28.}%
  \orphepsfiggy{#1}{#2}{#3}{#4}{#5}{#6}{#7}{#8}%
     \else%
\ifx\wherethehellami\nofigtex
  \noepsfiggy{#1}{#2}{#3}{#4}{#6}{#8}%
  \message{...(epsfiggy) 
   Assuming that this version of LaTeX does not permit insertion of}%
   \message{figures in EPS files.}%
   \message{(No "textures" or "orph" flag was specified.)}%
   \message{Writing caption and leaving space for figure,}%
   \message{but "#5" is being ignored.}%
\fi%
\fi%
\fi%
}%

\orph

\renewcommand{\arraystretch}{1.2}

\newcommand{\lefthead}{M. W. Guidry,  D. H. Feng, X.-W. Pan, and C.-L.
Wu}
\newcommand{\righthead}{Solution of the Nuclear
Shell Model by Symmetry-Dictated
Truncation}
\runningheads{\lefthead}{\righthead}

\renewcommand{\epsilon}{\varepsilon}
\renewcommand{\approx}{\simeq}
\renewcommand{\tightdot}{\kern-.2em\cdot\kern-.2em}

\newcommand{\heavyline}{\hline\noalign{\vskip 0.01em}\hline}

\title{\LARGE\bf \vspace*{-70pt}
Solution of the  Nuclear Shell Model by \\
Symmetry-Dictated Truncation\thanks{Review article prepared
for the {\em Journal of Physics {\bf G}.}}
 \vspace {15pt} }

\author {M. W.  Guidry,$^{1,2	}$   D. H.  Feng,$^3$
X.-W. Pan,$^{3}$ and C.-L. Wu,$^{4}$
\\ \hbox to 0pt{}           
\and
\utk{1}
\and
\ornl{2}
\and
\drexel{3}
\and
\taiwan{4}}

\smalldate

\begin{document}

\tableofcontents
\titlepage
\maketitle

\pagestyle{myheadings}
\thispagestyle{empty}

\begin{abstract}
\footnotesize\sl\noindent
\protect\baselineskip=20pt
The dynamical symmetries of the Fermion Dynamical Symmetry Model are
used as a principle of truncation for the spherical shell model.
Utilizing the usual principle of energy-dictated truncation to select a
valence space, and symmetry-dictated truncation to select a collective
subspace of that valence space, we are able to reduce the full shell
model space to one of manageable dimensions with modern supercomputers,
even for the heaviest nuclei.  The resulting shell model then consists
of diagonalizing an effective Hamiltonian within the restricted
subspace.  This theory is not confined to any symmetry limits, and
represents a full solution of the original shell model if the
appropriate effective interaction of the truncated space can be
determined.  As a first step in constructing that interaction, we
present an empirical determination of its matrix elements for the
collective subspace with no broken pairs in a representative set of
nuclei  with $130\le A \le 250$.  We demonstrate that this effective
interaction can be parameterized in terms of a few quantities varying
slowly with particle number, and is capable of describing a broad range
of low-energy observables for these nuclei.  Finally we give a brief
discussion of extending these methods to include a single broken
collective pair.

\end{abstract}


\newpage


\bibliographystyle  {prsty}      

\setlength{\baselineskip}{24pt}

\heading{introduction}{Introduction}

The shell model is fundamental to the understanding  of nuclear
structure, but it cannot be used for practical calculations in systems
with many valence particles outside closed shells.  There are two basic
reasons for this.  (1)~The matrix dimensionalities are too large.
(2)~There are too many effective interaction parameters.  The first
problem is well known; the second is as important, but is less
appreciated.  We may illustrate this second problem by noting that if
one views the effective interaction for a major shell of neutrons and
of protons as specified by a set of matrix elements to be determined by
fits to existing data, for the heaviest nuclei we must determine a
minimum of about 2500 parameters (matrix elements of allowed one and
two-body interactions) from some combination of theory and data.  By
comparison, the same approach in the $sd$ shell requires almost two
orders of magnitude fewer parameters.

\heading{2}{Symmetry as a Truncation Principle}

This proliferation of parameters is not a failure of the theory.  The
large number of parameters just reflects the number of independent one
and two-body scatterings that are possible within a large valence space
populated by significant numbers of nucleons.  Of course in all
practical calculations, the number of parameters has been reduced by
some means.  The most drastic prescriptions are the mean field ones
that replace these parameters by an average potential, and an
additional set of prescriptions (often in the form of semi-empirical
recipes) that prescribe how to calculate observable quantities, and how
to correct at some  level (again often semiempirical) for the
correlations neglected in such an approximation.

The applicability of mean field theories in a variety of applications
is well established, but the nuclear many-body problem cannot be
reduced to mean field terms, and the semiempirical nature of mean field
successes argues for a more microscopic understanding.  On the other
hand, the success of mean field methods, and the relative simplicity of
much of low-lying nuclear structure, suggest that most of the possible
contributions to the effective interaction either are negligible, or
more likely, they enter (low energy) nuclear structure only in certain
restricted combinations.

\subheading{energy-dictated}{Modern Approaches to the Shell Model Problem}

Thus, the solution of the general shell model problem for heavy nuclei
requires a resolution of the matrix dimensionality problem and a
consistent method to select a highly restricted subset of the effective
interaction parameters as the ones relevant for  low energy structure.
Ideally, one would like an approach that accomplishes both of these
tasks in a self-consistent manner.  Traditionally one begins with some
form of energy-dictated truncation.  This helps with the matrix
dimensionality problem by restricting the size of the shell model
space, and at the same time limits the number of parameters that must
be determined.  However, energy-dictated truncations have had limited
success for nuclei far removed from closed shells:  the influence of
any single high-lying configuration  may be negligible, but the
aggregate contribution of many such configurations
may not be.  Thus, energy-dictated truncations alone are
unlikely to allow us to solve the general shell model problem for heavy
nuclei.

There are four  modern approaches to this problem of trying to extend
the shell model to the description of heavy nuclei far removed from
closed shells.

\begin{enumerate}
\itemsep=-1pt

\item
Improved algorithms and computers for
traditional shell models (e.~g., Refs.\
\reference{val93,oxbash,pan95b}).

\item
Path integral solutions of the shell model using Monte Carlo algorithms on fast
supercomputers (e.~g., Refs.\ \reference{joh92,orm94}).

\item
Truncations of the shell model space based on guidance from mean-field
geometrical models with approximate pairing (e.~g., Refs.\
\reference{sch86,har91}).

\item
Symmetry-dictated truncations of the shell model space (e.~g., Refs.\
\reference{wu94b,gui93e,gui93g,iac87}).

\end{enumerate}

\noindent
Approaches (1) and (2) have had some success with the matrix
dimensionality problem, either by attacking it more efficiently, or by
avoiding it altogether in the path integral approach.  However, neither
offers an intrinsic solution to the burgeoning number of effective
interaction parameters in the heavy nuclei.  One  still must supplement
these approaches with a prescription for selecting the components of
the interaction that are to be emphasized.  We will not address methods
(1) or (2) in this review, and refer the reader to the literature cited
above on these subjects
\reference{val93,oxbash,pan95b,joh92,orm94}.

On the other hand, approaches (3) and (4) help with both aspects of the
problem:  the truncations implied by these prescriptions reduce the
matrix dimensionality, and the nature of the truncation suggests a
prescription for choosing the form of the most important interaction
terms in the resulting trunctated space.  An excellent example of
approach (3) is the Projected Shell Model \reference{har80,har91},
which truncates the space using a deformed mean field plus BCS pairing
prescription, and then diagonalizes a Hamiltonian composed of low-order
pairing and multipole terms in the truncated basis.  The approaches in
category (3) have had considerable success in practical calculations,
but we will not deal with them in this review, except to remark that
they have many deep similarities with the symmetry-based approaches
that we shall discuss.

We focus the remainder of our discussion on symmetry-dictated
truncation and its integrated solution to the shell model problem:
(1)~the symmetries dictate a severe truncation of the shell model
space;  (2)~the requirement that the dominant interactions respect
these symmetries provides a methodology for emphasizing a limited
subset of  interactions.  Whether such an integrated approach is useful
depends on the ability of the resulting theory to describe a broad
range of nuclear structure data   with a restricted set of parameters
that vary in a well-conditioned manner with particle number.

\subheading{symmetry}{Symmetry-Dictated Truncation}

Symmetry-dictated truncation is illustrated for a simple model assuming
an \su2 symmetry  in \fig{trun1}, and for a shell model dynamical
symmetry in \fig {trun2}.  Stated somewhat loosely, an energy-dictated
truncation truncates the space ``spherically'' in the space of symmetry
generators (for example, the three angular momentum components in
\fig{trun1}), but  a symmetry-dictated truncation reduces the space
further by selecting a particular ``direction'' (or set of directions)
in the space of symmetry generators.  Representative  directions are
indicated schematically in \fig{trun1} and \fig{trun2} by the heavy
arrows. Such an approach is closely associated with spontaneous
symmetry breaking and phase transitions because of the selection of
preferred directions in the space.

\subheading{effint}{Shell Model Effective Interactions}

Shell model effective interactions are highly dependent on the
truncation scheme \reference{kir85,wu87b,che91}.
  A  realistic initial approach to obtaining an effective interaction
is to treat  its matrix elements as parameters to be determined from
data\reference{tal52}.
The prototype of this method is the determination of the Brown
and Wildenthal matrix elements for light nuclei
\reference{wil85,bro88}, but a literal transcription of the empirical
$sd$ shell approach to the actinide nuclei  would require approximately
2500 matrix elements of the
 effective interaction to be determined from the data.

On the other hand, in the actinide region the most general $Sp_6^\nu
\times Sp_6^\pi$ Hamiltonian of the Fermion Dynamical Symmetry Model
\reference{wu94b} requires (at most) about 100 effective interaction
parameters to be determined, which is comparable to the number
 required in the full shell model for the $sd$ shell
 \reference{wil85}.  In reality, the symmetry-dictated truncation of
the {\small FDSM} will limit the number of relevant parameters even
more severely than these formal estimates would suggest:  the results
to be  presented here indicate that most low-lying nuclear
properties are determined by only of order 10 symmetry-selected
parameters for the entire shell, with (at most) a weak dependence of
these parameters on particle number.

\subheading{effint-fdsm}{FDSM Effective Interactions}

Thus, the  {\small FDSM}  symmetry-dictated truncation provides a
methodology with the potential to enable shell model calculations for
all nuclei. To establish its validity, it is necessary to demonstrate
through a detailed set of calculations that the required effective
interaction (1)~is  easily determined, and  (2)~produces a quantitative
description of observables in heavy nuclei.  This  will be an iterative
process requiring a long series of systematic calculations, but we
offer the following initial speculations  concerning the  {\small
FDSM}  effective interactions and the process of determining  it:

\begin{enumerate}
\itemsep=-1pt

\item
The symmetry-dictated truncation of the {\small FDSM}  is qualitatively
different from energy-dictated truncations; it could lead to effective
interactions  in the highly truncated spaces that are markedly
different in the components that are emphasized relative  to the
appropriate interactions in the full shell model space.  Unlike
energy-dictated truncations, a  symmetry-dictated truncation selects
particular ``directions'' in the shell  model space (see \fig{trun2}).

\item
A complicated particle number dependence for the effective interaction
would make the theory difficult to use in a consistent microscopic
way.  Therefore, we require that the effective interaction have a
smooth and/or weak particle number dependence within a shell.  Since
filling different valence shells defines different phases of the
theory, it is possible (but not required) that the parameters could
have discontinuities between shells.  Likewise, it is possible (but not
required) that there may be discontinuities between parameter sets
describing regions having different dynamical symmetries, since these
too define different phases of the theory.

\item
A reasonable first step will be to determine the effective interaction
in the $SD$ collective subspace of the {\small FDSM} using
symmetry-limit calculations.  We may then investigate deviations from
the symmetry limits within the $SD$ space,  and may expect that this
leads to small and smooth changes of the  effective interaction
parameters from the symmetry-limit values.

\item
Finally, we may expand the space to include broken collective pairs.
This will allow the determination of additional effective interaction
parameters associated with high-spin physics that played small roles in
the low-lying states.  In addition,  we may expect a systematic
renormalization of the  parameters already determined in the $SD$
subspace by virtue of enlarging the space.  This renormalization under
space enlargement must be well behaved if the
symmetry-dictated method of truncation is to be of practical utility.

       \end{enumerate}

\noindent
Thus an important first step in this program is to take dynamical
symmetry limits as a starting point and fit to the data using an
{\small FDSM}  Hamiltonian with symmetry breaking terms included in
lowest order perturbation around the symmetry limits.  Applications of
the {\small FDSM} at this level have been discussed extensively in
\reference{wu94b}.  At the next level of improvement, we may assume the
most general highest symmetries and a corresponding Hamiltonian, but
not restrict the calculation to the dynamical symmetry limits or to
perturbation around the symmetry limits.
 We may hope to obtain by these approximations a qualitative agreement
with experiment; this effective interaction can then be refined in
 subsequent calculations that include more realistic configuration
 mixing and enlarge the space to include broken pairs.

In this review we shall emphasize two aspects of this extension of the
{\small FDSM}
 beyond the dynamical symmetry limits.  The first is the initiation of
 a systematic analysis of symmetry-breaking terms in the realistic
shell model Hamiltonian relative to the {\small FDSM} symmetry limit
Hamiltonians.  The second is a series of numerical calculations that
take the theory beyond the symmetry limits and begin to establish a set
of effective interaction parameters for realistic symmetry-truncated
shell model calculations.

\heading{3}{Examples of Truncated Valence Spaces}

The essence of the {\small FDSM}
 approach is to impose a symmetry-dictated truncation within a
particular valence space, with the choice of valence space often
dictated by energy considerations.  As has been discussed in some
detail in Ref.\ \reference{wu94b}, there is considerable flexibility in
the choice of the space in which to implement such a truncation.  The
space truncation is instituted through the generalized Ginocchio
\reference{gin80} coupling scheme illustrated on the far right side of
 \fig{belyaev5} \reference{wu94b}.  The most common implementation has
been within a single major shell, as illustrated by the coupling
labeled ``Normal Deformation'' in \fig{belyaev5}.   However, an {\small
FDSM} model of superdeformation has also been proposed \reference{wu92b}
that uses the alternative coupling scheme labeled ``Super Deformation''
in  \fig{belyaev5}.   This example makes it clear that the choice of
valence space is not the central issue, but a clever choice of the
valence space may enhance the ability of the symmetry limits of the
theory to describe data.

More generally, it has been suggested that the {\small FDSM}
 can be formulated in a variety of valence spaces, with different
choices associated with what would be termed shape coexistence in a
mean-field theory \reference{gui93d}.  In this review we shall
concentrate on applications of symmetry-dictated truncation to
major-shell valence spaces, but the basic ideas should be applicable to
a broad range of valence spaces if one can implement  a
symmetry-dictated truncation within them.  In particular, we expect
that a similar analysis is possible for superdeformed states.

\heading{4.2}{Symmetry-Breaking in the FDSM}

   The symmetry limits of the {\small FDSM}, and perturbation theory
about those symmetry limits, describe a broad range of nuclear
structure observations \reference{wu94b}.  It is of interest to
examine  systematically the excursions from the symmetry limits of the
theory, in order to test its suitability as a systematic truncation
procedure for quantitative shell model calculations in heavy nuclei.

\subheading{breaking}{The Breaking of Dynamical Symmetries}

Principles of dynamical symmetry and dynamical symmetry breaking are
summarized in \fig{belyaev8}.  A dynamical symmetry results when a
Hamiltonian can be expressed in terms of invariants for the highest
group and subgroups of a group chain, as we illustrate for the $Sp_6
\supset SU_3\supset SO_3$ dynamical symmetry of the {\small FDSM}.
Symmetry-breaking then corresponds to the presence of
Hamiltonian terms that cannot be expressed in this
 manner.  Figure \ref{fg:belyaev8} also distinguishes two approaches to
dynamical symmetry breaking.  The first is empirical:  one defines the
symmetry breaking to be the difference between the observation and the
symmetry-limit calculation.  This provides a definition of symmetry
breaking, but is not predictive.  More useful are the microscopic
approaches to symmetry breaking,  where the symmetry-limit Hamiltonian
$H\tsub{sym}$ is presumed to  derive from a more fundamental
Hamiltonian $H\tsub{micro}$ with a known form.  Therefore, one can use
theoretical and physical arguments to identify likely symmetry breaking
terms $H\tsub{sb}$, and can make {\em predictive} estimates for the
expected magnitude of symmetry breaking.  The situation addressed here
falls in this second category:   the {\small FDSM} symmetry-limit
Hamiltonian represents an approximation to the full shell model
Hamiltonian that omits terms breaking a particular dynamical symmetry.
Since the forms of the two Hamiltonians are known, we may use physical
arguments to identify the terms of the full Hamiltonian that break the
symmetry and are likely to be important.  We now illustrate
using $\Sp6 \supset \su3$ symmetry breaking by spherical
single-particle energies in the {\small FDSM} \reference{wu94a}.

\subheading{single-particle}{Symmetry Breaking by
Spherical Single-Particle Energies}

Orbitals exhibiting an $Sp(6) \supset SU(3)$ dynamical symmetry have
$k=1$  (see \fig{belyaev5}).  A Hamiltonian with an $Sp(6) \supset
SU(3)$ dynamical symmetry requires degeneracy in the single-particle
energy terms corresponding to the same value of $i$.  For normal
deformation in heavier nuclei, and for superdeformation, there are
multiple values of $i$ within a valence shell and  the symmetry-limit
Hamiltonian will exhibit a higher level of degeneracy than the
realistic spherical single-particle shell model spectrum
(\fig{belyaev5}).  The difference between the symmetry-limit and
realistic spectra of \fig{belyaev5} are shown in \fig{belyaev6}, which
also summarizes several issues that are relevant to the present
discussion.  These concern, not simply the size of the single-particle
splitting, but its size relative to the correlation energy of the
system and  how much of that splitting breaks the relevant symmetry
\reference{wu92b,wu94b,gui93d}.

The states of the {\small FDSM} are classified according to a total
heritage quantum number $u$ that measures the number of particles not
coupled to coherent $S$ and $D$ pairs \reference{wu87b}.  Low-spin
states of even--even nuclei are dominantly $u=0$ configurations.  The
mixing matrix elements associated with the splitting of the
single-particle energies may be expressed as
\begin{equation}
\left< \lambda' \mu' u' \left| n_i^{(rr)0} \right| \lambda \mu u \right> =
\frac{\left< \lambda' \mu' u' \left| [n_i^{(rr)0}, C_{SU(3)}] \right|
\lambda \mu u \right>} {C(\lambda\mu) - C(\lambda'\mu')}.
\end{equation}
where $u$ and $u'$ are the heritage quantum numbers,
$C(\lambda \mu)$ is the usual eigenvalue of the quadratic $SU_3$ Casimir
operator $C_{SU(3)}$ evaluated in an $SU_3$ representation $(\lambda, \mu)$,
and
the single-particle energy expressed in terms of the standard {\small FDSM}
$k$--$i$ basis is:
\begin{equation}
\sum_{j} n_{j}e_{j}=\sum_{r, i} { n}^{(rr)0}_{i} e^{r}_{i},
\label{1}
\end{equation}
  where the operator ${ n}^{(rr)0}_{i}$ is given by
\begin{equation} { n}^{(rr)0}_{i} =\sqrt{2\Omega_{i}} \left[ b_{ki}^{\dagger}
\tilde{b}^{\phantom{\dagger}}_{ki} \right]^{(rr)0},
\label{2}
\end{equation}
and
\begin{equation}
e^{r}_{i}=\sum_{j}e_{j}
\left[ \begin{array}{ccc} k&i&j\\  k&i&j\\ r&r&0 \end{array} \right] \sqrt{
\Omega_{j} / \Omega_{i}}.
\label{3}
\end{equation}
with the square bracket denoting a normalized 9-j  coefficient.  The
$k$--$i$ basis $b_{ki}^{\dagger}$ has been defined in
\reference{gin80,wu86,wu87b}; $\Omega_{j}$ and $\Omega_{i}$ are the
pair degeneracies for the $j$ shell and the shells associated with
pseudospin $i$, respectively [$\Omega_{j}=j+\frac{1}{2}$, and
$\Omega_{i}=(2k+1)(2i+1)/2$].  Ref.\ \reference{wu94a},
demonstrates that the only  operator capable of mixing an $SU_3$ irrep
in the $u=0$ bands  with other $SU_3$ irreps is ${ n}^{(11)0}_{i}$.
Estimates for the upper limit on the symmetry breaking associated with
this term are summarized in Fig.\ \ref{fg:belyaev1} for typical heavy,
normally-deformed  nuclei.  A similar analysis for superdeformation
suggests that the single-particle symmetry breaking associated with the
{\small FDSM} model of superdeformation is even smaller than that
exhibited in this example for normal deformation.  Thus, we expect that
spherical single-particle symmetry breaking in both normal and
superdeformed nuclei will be perturbative in size for the $\Sp6 \supset
\su3$ symmetry of the {\small FDSM}, and the full inclusion of such
terms will be unlikely to invalidate previous symmetry-limit results.

\subheading{size}{Comments on the Size of Symmetry Breaking Terms}

The stability of the $SU_3$ dynamical symmetry for the {\small FDSM}
 derives from the particular structure of the Ginocchio $S$--$D$ pairs
\reference{gin80,wu87b,wu94b}.  The single-particle terms can break the
$SU_3$ symmetry only through an indirect Pauli effect associated with
the embedding of $SU_3$ in the higher symmetry $Sp_6$. The only
operator in the single-particle
 energy terms that can accomplish this is ${ n}^{(11)0}_{i}$.  In other
fermion theories, such as the Elliott model
 \reference{ell58} or the pseudo-$SU_3$ model \reference{hec69,ari69},
 the $SU_3$ symmetry is   embedded in a much larger group and there are
many generators that could  break the symmetry directly.

This illustrates an important concept concerning symmetry breaking:
the size of the symmetry-breaking terms for a particular dynamical
symmetry may differ considerably from that expected based on experience
with mean field theories, or even theories based on a symmetry that is
formally related but  implemented in a different physical manner.  In
this case we see that the single-particle terms have little influence
on the \su3 symmetry because of the unique properties of the pairs that
generate the symmetry.  On the other hand, in a deformed mean field theory one
generally finds that the quantitative results are much more sensitive
to the details of the initial spherical single-particle spectrum.

\heading{5}{Heritage-0 Calculations for Major Shell Truncation}

Let us now examine some {\small FDSM} calculations.  The majority of
these will employ  the code {\small FDU0} of Wu and Vallieres
\cite{hwu89,val91} that diagonalizes the most general interaction
consistent with the highest shell symmetry of the valence space.  Thus,
its solutions may be considered to be linear combinations of the
dynamical symmetries allowed in a valence space.  It is valid only for
heritage-0 subspaces;  thus,  it is most applicable for states in
even--even nuclei below angular momentum 10, where the breaking of
pairs is not too important. In addition, this code assumes that $N_1$,
the number of pairs in the normal parity orbitals of the valence shell
(for neutrons or protons), is conserved.  Thus, it does not incorporate
directly the effects of scattering particles between the normal and
abnormal parity orbitals (such effects may still be included indirectly
in the effective interaction \reference{li94a,gui93f}).
Because this scattering is expected to
be more important in the Pauli forbidden region lying between 1/3 and
2/3 filling of the normal parity orbitals in \Sp6 shells
\reference{wu94b,gui93c}, the validity of {\small FDU0} in that region
is also questionable.  Therefore, in this review we shall confine
attention to states of low angular momentum  for nuclei that do not lie
in the Pauli forbidden region of \Sp6 shells.  For \so8 shells, there
is no such restriction and we are free to apply {\small FDU0}
calculations to all nuclei in the shell.

\subheading{hamiltonian}{Hamiltonian}

The most general Hamiltonian consistent with
an \so8 or \Sp6 highest shell symmetry may be expressed as
\cite{hwu89,wu94b}
   \begin{eqnarray}
       H \align= \sum _{\sigma=\pi, \nu}
       \left( B_2^\sigma P^2_\sigma
       \tightdot P^2_\sigma + B_3^\sigma P_3^\sigma \tightdot
        P_3^\sigma
       +G_0^\sigma S^\dagger_\sigma S\pdagger_\sigma
       + G_2^\sigma D^\dagger_\sigma D\pdagger_\sigma \right)
     \nonumber
\\
       \flush + B_2^{\pi\nu}P_2^\pi \tightdot P_2^\nu
       + B_3^{\pi\nu} P_3^\pi \tightdot P_3^\nu
       + B_1^\nu P_1^\nu \tightdot P_1^\nu
       + B_1^\pi P_1^\pi \tightdot P_1^\pi
       + B_1^{\pi\nu}P_1^\nu \tightdot P_1^\pi
       \tag {fduo}
   \end{eqnarray}
where $S$ denotes monopole pair operators, $D$ denotes quadrupole pair
operators, and $P_r$ denotes multipole operators of order $r$, with all
quantities constructed in the $k$--$i$ truncation scheme illustrated in
\fig{belyaev5}.
Not all  of these terms contribute for a particular highest symmetry.
One finds that for
spectra (which depend only on energy differences) there are
at most 11 parameters for  $SO_8^\pi \times SO_8^\nu$, 8 for $Sp_6^\pi
\times Sp_6^\nu$, and 9 for $SO_8^\pi \times Sp_6^\nu$.  In the
numerical calculations that follow, we will typically use a simplified
version of this Hamiltonian that retains only pairing and quadrupole
terms, thereby reducing the number of free parameters to 5 or less for most
cases.

\subheading{n1}{Particle Number Distribution}

In the calculations to be presented in this review, the
number of pairs $N_1$ in the normal-parity levels is treated as a good
quantum number and is estimated from the semi-empirical formula determined
globally from the ground state spin of the odd-mass nuclei \reference{wu94b}:
\begin{equation}
    N_{1}=\left\{ \begin{array}{ll}
               N &~ \mbox{for}~ N < 1.5    \\
     0.75 + 0.5N &~ \mbox{for}~ 1.5 < N < 2\Omega_{0} + 1.5   \\
  N-\Omega_{0}   &~ \mbox{for}~ N > 2\Omega_{0} + 1.5 \end{array}
       \right.\tag{n1dist}
\end{equation}
where $N$ is the number of valence pairs and $\Omega_{0}$ is the pair
degeneracy of the abnormal-parity level. The above formulas are
suitable for both protons and neutrons, and are found to be in
excellent agreement with a similar distribution obtained empirically
from the Nilsson level scheme with a measured deformation parameter
$\beta$.

\subheading{5.1.2}{Masses of Heavy and Superheavy Nuclei: Z=110--111}

In recent years, evidence has  accumulated
\reference{mun81,mun82,mun84} for extension of the known elements to
larger proton number, culminating in the discovery of
 several isotopes having proton number $Z=110-111$ \cite{hof95}.  The
usual view
 is that these isotopes represent the tail of the ``normal element''
distribution, and that the predicted superheavy elements represent a
qualitatively different set of nuclides that remain undiscovered at a
predicted proton number  of about 114 and a neutron number
approximately 20 units heavier than for the recently-discovered
isotopes of elements 110 and 111.  As we now discuss, truncated shell
model calculations suggest a different interpretation of these results
that could have important experimental implications
\reference{han92,gui95a}.

\subsubheading{mass}{Comparison of Mass Calculations for Very Heavy Elements}

In \tableref{sheavy}, we compare the observed mass excesses (first row,
labeled ``Exp'') for isotopes of the heaviest elements with
calculations for these masses.  A perusal of these results suggests
that  the {\small FDSM} mass calculations described in Ref.\ \reference{han92}
give a global description of the heaviest elements that is arguably the
best now available.  This is particularly noteworthy because the {\small FDSM}
calculations were {\em not} tuned specifically to these heaviest
elements;  the parameters employed in Ref.\ \reference{han92} were
determined by a global fit to all available masses above $Z=82$ and
$N=126$.  We also emphasize that the accurate description of the
$Z=110-111$ isotopes represents a pure prediction of the {\small FDSM} theory,
published before their discovery, and  without adjustments to the
theory based on specific properties of the $Z=108-109$ elements that
were known at the time of the calculations.

\subsubheading{superheavy}{Superheavy Elements}

An island of superheavy elements was found in the {\small FDSM}
 mass calculations \reference{han92}, but this island was shifted to
considerably lower neutron number than is predicted in classical
calculations of the stability of superheavy elements, was found to be
more stable than in such calculations, and was found to correspond to
nuclei having near-spherical shapes.  The shell correction associated
with this minimum is illustrated in \fig{esurface} for the case of a
spherical Woods--Saxon single-particle spectrum.  The location of the
new predicted maximal shell stabilization,  the traditional location of
the island of superheavy stability, and the new isotopes of elements
110--111, are also shown in \fig{esurface}.  These properties of the
{\small FDSM} heavy-element solutions are consequences of two general
physical principles.  The first is a monopole--monopole interaction
that is expected  to be present on general shell-model grounds, and
found from empirical mass fits to become increasingly repulsive in
moving away from the stable nuclei.  The second is that the structure
of these heaviest elements is  expected to involve a competition
between an $SU_2$ dynamical symmetry favoring spherical symmetry and an
$SU_3$ dynamical symmetry favoring axially symmetric deformation
\cite{wu94b}.

\subsubheading{experimental}{Experimental Implications}

Since Ref.\ \reference {han92} was a general treatment of nuclear
masses that was not tuned to predict the properties of these heavy new
isotopes, we are emboldened to draw some general conclusions from this
successful prediction that  differ from those of more traditional
approaches to the stability of the heavy elements \reference{gui95a}.

\begin{enumerate}
\itemsep= -1pt
 \item
The observation of  isotopes with proton number $Z=110-111$ having the
predicted properties  may indicate a shift  of the superheavy island of
stability  to considerably lower neutron numbers, as predicted in Ref.\
\cite{han92}.

\item
These new isotopes are expected to be near-spherical or
deformation-soft  structures, in contrast to the heavy elements with
$Z\le 106$, which are found in the {\small FDSM} (and  other)
calculations to have well-deformed ground states, and to most other
descriptions of the new $Z=110-111$ elements, which find them to have
deformed ground states.

\item
The successful prediction of the masses of these new isotopes is
further evidence supporting the importance
of the monopole--monopole interaction
introduced in \cite{han92} for the description of nuclear masses.

\item
No shell stabilization minimum is expected at the traditional location
of the superheavy elements because of the repulsive monopole--monopole
interaction.  Thus, we predict that the traditional island of
superheavy elements does not exist,  and that the r-process element
production path may run closer to the stability valley than has
generally been assumed for the heaviest elements \cite{wan92}.
\end{enumerate}

\noindent
The present hypotheses have some consequences that are testable, though
the experiments are difficult.  (1)~There should exist  nuclides of
elements $Z=$ 112--114 that are approximately as stable as those of elements
110--111 (see \fig{esurface}).  (2)~Beyond $Z\approx 116$ and $N\approx
170$, the heavy nuclei should rapidly become less stable, and the
region of traditional superheavy nuclei should be quite unstable (see
Fig.\ \ref{fg:esurface}).  (3)~The nuclides in this new region of
superheavy nuclei ($Z\approx 110-116$ and $N\approx 160-170$) are
expected to be either spherical or very deformation soft, with
attendant consequences for properties such as the alpha decay
systematics.  (4)~The implied shift of the r-process path closer to the
stability valley should also have observable consequences, but this may
require an improved understanding of the astrophysical environment for
the r-process.

\subheading{5.2}{The Xe--Ba Region}

The {\small FDSM} is defined by a fermionic Lie algebra, has symmetry
limits analogous to all the {\small IBM} limits, and takes the Pauli
principle into account \reference{gin80,che86}. Furthermore,  the
states for even and odd systems in the {\small FDSM} belong to vector
and spinor representations, respectively, of $SO_8$ or $Sp_6$, thus
allowing a description of even--even and even--odd nuclei  without  the
necessity of additional degrees of freedom.  Therefore,
symmetry-truncated shell model calculations based on the {\small FDSM}
dynamical symmetries for even and odd nuclei offer a microscopic
alternative to {\small IBM} and nuclear supersymmetry ({\small NSUSY})
as means to unify the structure of even and odd systems.

Nuclei in the Xe--Ba region have neutrons and protons in the 6th shell,
with {\small FDSM}
 pseudo-orbital angular momentum $k=2$ and pseudo-spin $i=\frac{3}{2}$
for the normal-parity levels.  Thus they are expected to possess
$SO^{\pi}_8\times SO^{\nu}_8$ symmetry, which contains an
$SO_8^{\pi+\nu}$ subgroup, and to have analytic solutions for $SO_5
\times SU_2$, $SO_6$, and $SO_7$ dynamical symmetries.  For these
reasons, the Xe--Ba region is an excellent one in which to investigate
a unified {\small FDSM} description of even--even and even--odd nuclei.

\subsubheading{5.2.1}{Energy Spectra}

The wavefunctions for both even and odd nuclei are given by the
following group chain
\begin{eqnarray}
  &(~SO^{i}_{8}~\supset~SO^{i}_{6}~\supset~ SO^{i}_{5}~)
 \times SO^{k}_{5}~~
  \supset~ SO^{i+k}_{5}~ \supset SO^{k+i}_{3}~~~~~~~~~~~~~~~ & \\
  &~~[l_{1}l_{2}l_{3}l_{4}] \hspace {0.4cm}
 [\sigma_{1} \sigma_{2} \sigma_{3}]\hspace{0.4cm}
  [\tau_{1}\tau_{2}] \hspace{0.7cm} [\tau ]\hspace{1.2cm}
  [\omega_{1}\omega_{2}]\hspace{0.8cm}
  J~~~~~~~~~~~~~~~~~~~~~~~&  \nonumber
\label{eq:eq21}
\end{eqnarray}
where $[l_{1}l_{2}l_{3}l_{4}]$, $[\sigma_{1}\sigma_{2}\sigma_{3}]$,
and $[\tau_{1}\tau_{2}]$ are the Cartan--Weyl labels for
the groups $SO_8$, $SO_6$, and $SO_5$, respectively, $\tau =0 (1)$
for even (odd) nuclei, and $k$ and $i$ indicate pseudo-orbital and
pseudo-spin parts of the groups, respectively.
The energies for even systems are
\begin{eqnarray}
E\tsup{even} = E^{\scriptsize\rm
(e)}_0 + g_6 \sigma (\sigma + 4) + g'_5 \tau (\tau +3)
+ g'_I J (J+1)  ,
\label{eq29a}
\end{eqnarray}
and for odd systems,
\begin{eqnarray}
&E\tsup{odd} =& E^{\scriptsize\rm
(o)}_0 + g_6 [\sigma_1 (\sigma_1 + 4) + \sigma_2
(\sigma_2 + 2) + (\sigma_3)^2] + g_J J(J+1)
\nonumber\\
&&+ (g'_5 - g_5) [\tau_1
(\tau_1 +3) + \tau_2 (\tau_2 + 1)] + g_5 [\omega_1 (\omega_1 + 3) +
\omega_2 (\omega_2 + 1)] .
\label{eq29b}
\end{eqnarray}
More details on the construction of these formulas, and the
reduction rules for even and odd systems, are given in
Ref.\ \reference{pan95a}.

The low-energy spectra for $^{120-132}$Xe isotopes predicted by
Eq.\ (\ref{eq29a}) are compared with data in \fig{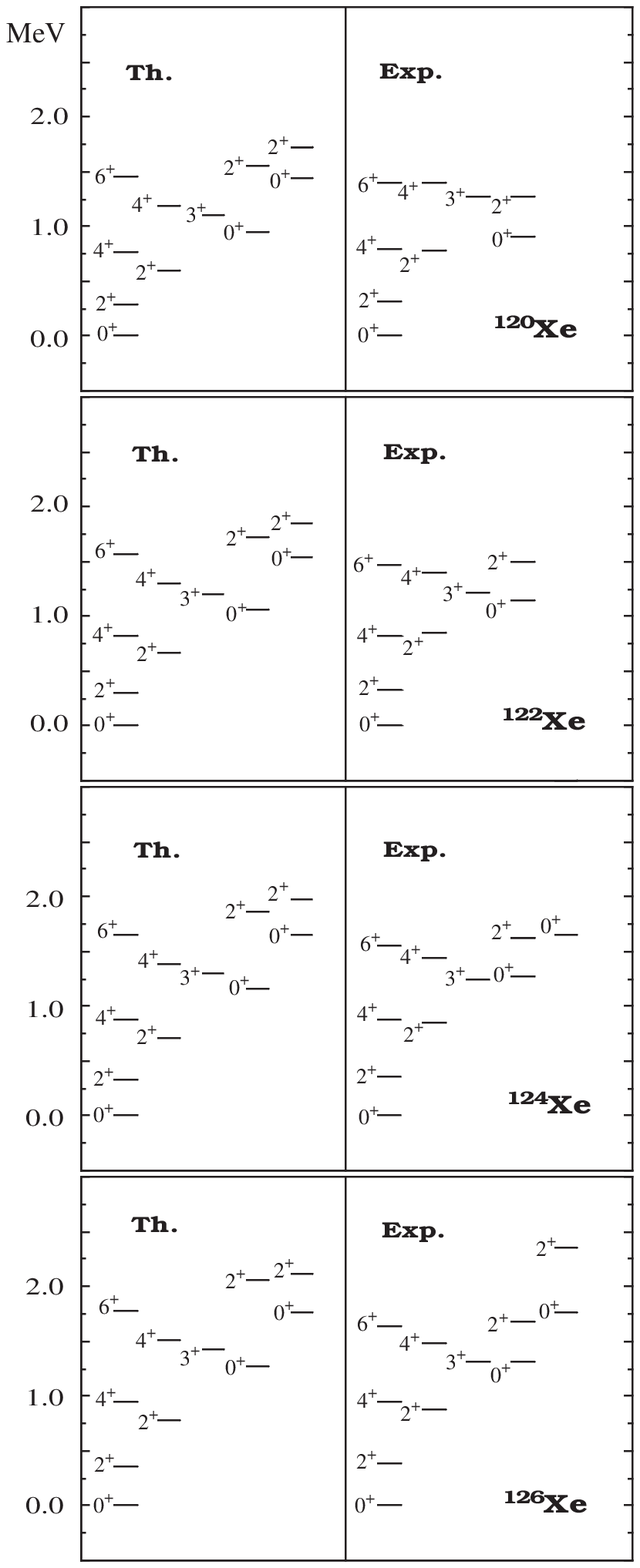} and
\fig{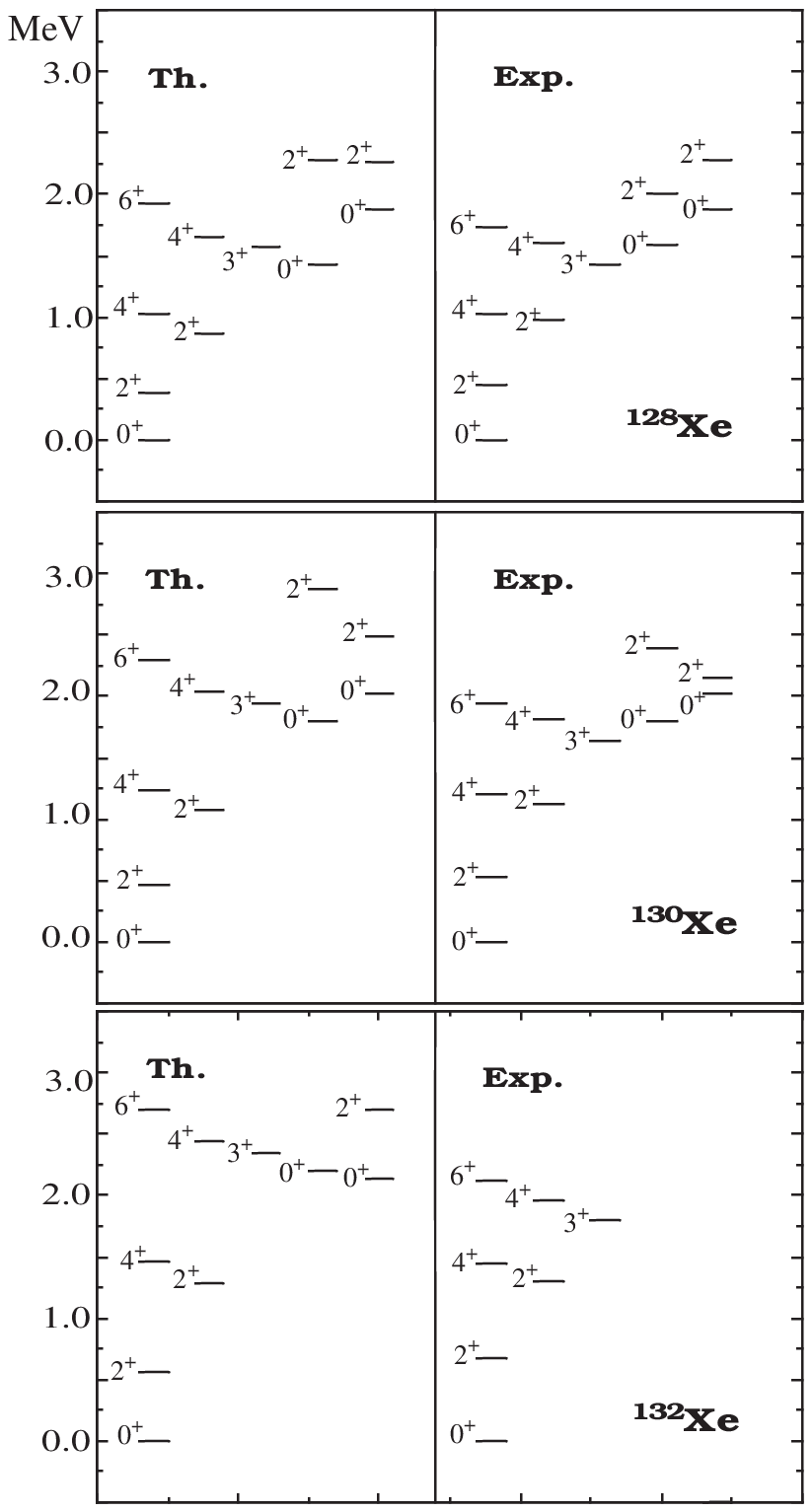}, and the parameters used in the calculations are given
in \tableref{xe1}. The experimental spectra indicate that the $SO_{3}$
parameter $g'_{I}$ is not sensitive to the neutron number in fitting
the spectra of a chain of isotopes (including both even--even and
even--odd nuclei).  Therefore, in fitting the even--even nuclear
spectra we fix the $g'_{I}$ parameter to be 11.9 keV and the adjustable
parameters were taken to be
 $g_{6}$ and $g'_{5}$.

Using \tableref{xe1} and Eq.\ (\ref{eq29b}), the spectra of the
neighboring even--odd Xe isotopes can be calculated.
We
present the calculated and experimental results for $^{127-133}$Xe in
Figs.\ \ref{fg:xeba3.eps}--\ref{fg:xeba4.eps}, with the parameters
given in \tableref{xe2}. Similar results for Ba isotopes may be found
in Ref.\ \reference{pan95a}.  Apart from the constant term, the formula
for $E\tsup{even}$ contains three parameters, and that for
$E\tsup{odd}$ contains two parameters beyond the three parameters that
are determined by fitting the spectra of neighboring even--even
nuclei.  With two extra parameters, Eq.\ (\ref{eq29b}) reproduces
the spectral patterns for the nuclei $^{127-133}$Xe and $^{131-135}$Ba
(not shown here; see Ref.\ \reference{pan95a}) with about 15 levels
each.

\subsubheading{5.2.3}{Electromagnetic Transition Rates}

In Ref.\ \reference{pan95a}, formulas are presented for the reduced
quadrupole transition strengths $B(E2)$ in both even and odd systems.
Table \ref{xebe2} summarizes some $B(E2)$ values calculated for
even--even Xe isotopes.  Similar results for Ba isotopes may be found in Ref.\
\reference{pan95a}.  These are seen to agree rather well with the
observations, and are indicative of a good \so6 symmetry in
this region.  In Table \ref{xe129-130}, both experimental and
theoretical
 $B(E2)$ values for the even--odd nuclei of $^{129}$Xe and $^{131}$Xe
are given, and compared with the results of  {\small
NSUSY} calculations.  The agreement with data is comparable in the two cases.

\subsubheading{5.2.4}{Fermion Dynamical Symmetry versus Supersymmetry}

In Ref.\ \reference{pan95a} it is shown that when the $u=1$ fermion
\so6 state for the {\small FDSM} core is replaced by a boson state and
the Pauli factor is ignored, the {\small FDSM} wavefunction goes over
to an {\small NSUSY} wavefunction.  Thus, {\small NSUSY} can be
obtained as an approximation to the {\small FDSM} for odd-mass $SO_6$
nuclei.  The full {\small FDSM} without this approximation implies
corrections to the {\small NSUSY} picture.

In the $U(6/4)$ {\small NSUSY}, the ground-state representation is
$[{1\over 2} {1\over 2}]$ and the ground-state spin is always ${3\over
2}$. However, both ${1\over 2}$ and ${3\over 2}$ are observed as
ground-state spins in this region. The second $SO_{5}$ Casimir operator
in Eq.\ (\ref{eq29b}) allows this possibility: for alternative signs of
the parameter $g_{5}$, the ground state spin can take the values
${1\over 2}$ or ${3\over 2}$ .  What is more, by allowing $g_{5}$ to
change smoothly from positive to negative, the systematic shift of the
ground band for the Xe (and Ba) isotopes can be reproduced, as shown
for Xe in \fig{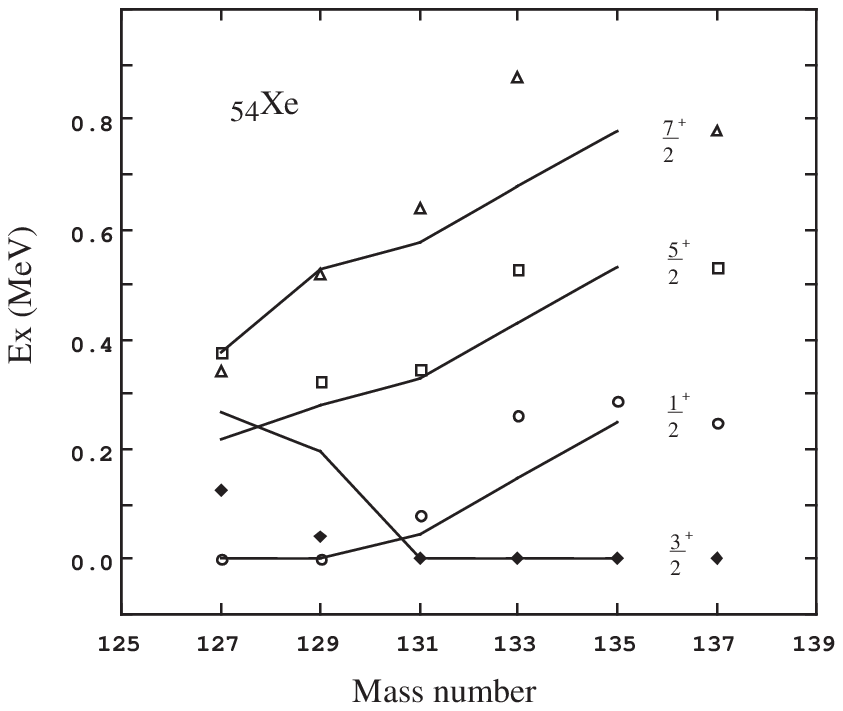}.

The general conclusion of Ref.\ \reference{pan95a} is that the {\small
FDSM} symmetry-dictated solution in the mass-130 region provides a
unified description of even and odd nuclei that is comparable to, or
better than, {\small IBM} and {\small NSUSY} approaches, but with a
firmer microscopic basis.  As one example of this microscopic basis,
the group chain corresponding to odd-mass \so6 nuclei in the {\small
FDSM}  is very similar to the corresponding {\small NSUSY} group chain,
but  pseudo-orbital angular momentum 2 is an {\it ad hoc} introduction
in the {\small NSUSY}, while in the {\small FDSM} it is a natural
result of the $k$--$i$ reclassification for the shell model
single-particle states in  the sixth major shell.

\subsubheading{5.2.5}{Tau Compression}

Pure $SO_{6}$ spectra give reasonable agreement with  data for $\tau
\leq 3$  states in $^{120-126}$Xe (see
Figs.\ \ref{fg:xeba1.eps}--\ref{fg:xeba2.eps}).  However, the
experimental energies for the higher $\tau$ values in $^{128-132}$Xe
are lower than predicted, with the discrepancies increasing with
increasing $\tau$.  Energy levels within the same $SO_5$ irrep (same
$\tau$) follow the $J(J+1)$ rule rather well, so this discrepancy
cannot be caused by the usual stretching effect.  In
Ref.\ \reference{pan94}, it was shown that this is an $SO_{5}$
$\tau$-compression effect whose driving force  is the reduction of
pairing correlation with increasing $\tau$.  Allowing for $g_{S} =
G_{0} - G_{2} \not= 0$, thereby deviating from the $SO_{6}$ limit, and
treating the $g_{s}S^{\dagger}\cdot S$ term as a perturbation leads to
the following energy formula, \begin{eqnarray} E'\tsub{even} \cong
E^{\scriptsize\rm (e)}_{0} + g_{6}\sigma (\sigma +4) +  A'\tau (\tau
+3) - B' [\tau (\tau +3)]^{2} +  g'_{I} J (J+1) .  \label{eq216}
\end{eqnarray} Fig.\ \ref{fg:xeba8.eps}
 shows the spectrum for $^{126}$Xe predicted by Eqns.\ (\ref{eq29a})
and (\ref{eq216}), respectively.  Inclusion of the $SO_{5}$ stretching
effect improves the agreement with data
significantly. Details and more examples
may be found in Ref.\ \reference{pan94}.

\subheading{5.3.1}{SO(6)-Like Behavior in the Pt Region}

Numerical studies using the {\small FDSM} have shown that there is an
effective $SO(6)$ fermion dynamical symmetry in the  \isotope{196}{Pt} mass
 region \reference{fen93,pin95}.  More recently, a systematic
description of nuclear structure has been given in this region using
symmetry-truncated shell model methods \reference{pin95}.  In this
section, we review these calculations for the platinum isotopes and the
emergence of effective $SO(6)$-like behavior from numerical
calculations near $^{196}$Pt, where the {\small FDSM} has no formal
\so6 symmetry.

\subsubheading{pthamil}{Hamiltonian and Spectra}

The five-parameter  Hamiltonian
is taken to be
\begin{equation}
 H=G'_{0\pi}S^{\dag}_{\pi}S_{\pi}^{\phantom{\dag}}+
    G'_{0\nu}S^{\dag}_{\nu}S_{\nu}^{\phantom{\dag}}+
    B'_{2\pi} P^{2}_{\pi} \cdot P^{2}_{\pi}+
    B'_{2\nu} P^{2}_{\nu} \cdot P^{2}_{\nu}+
    B_{2\pi\nu} P^{2}_{\pi} \cdot P^{2}_{\nu} .
\end{equation}
In the $SO(8)$ and $Sp(6)$ algebras, the quadrupole pairing interaction
in this Hamiltonian is not  independent  and can be accounted for by
redefining the parameters $ G'_{0\sigma}=G_{0\sigma}-G_{2\sigma}$ and
$B'_{r\sigma}=B_{r\sigma}-G_{2\sigma}$ ($\sigma=\pi,\nu$).  The spectra
of five even platinum isotopes $^{190-198}$Pt (there are approximately
13 levels for each nucleus) are found to be satisfactorily described by
sets of five parameters. In Figs.\ \ref{fg:pt.fig2.eps} and
\ref{fg:pt.fig1.eps} the calculated spectra for $^{190-198}$Pt are
compared with the data; the  corresponding parameters are summarized in
\tableref{ptparams}.

Fig.\ \ref{fg:pt.fig1.eps} indicates that most low-lying  $0^{+}$ and
$2^{+}$ states are well reproduced by the {\small FDSM}
 calculations,  tolerable agreement is obtained for the $4^{+}$ states,
and there is a persistent deviation of the calculated $3^{+}$ states
that increases with mass. The basic theoretical trends are the
narrowing of the energy gap between the $3^{+}_{1}$ and $4^{+}_{1}$
levels, and a widening of the gap between the $3^{+}_{1}$ and
$2^{+}_{2}$ levels.  This level pattern is characteristic of the \so6
dynamical symmetry and appears  naturally in the heavier Pt isotopes.
In \fig{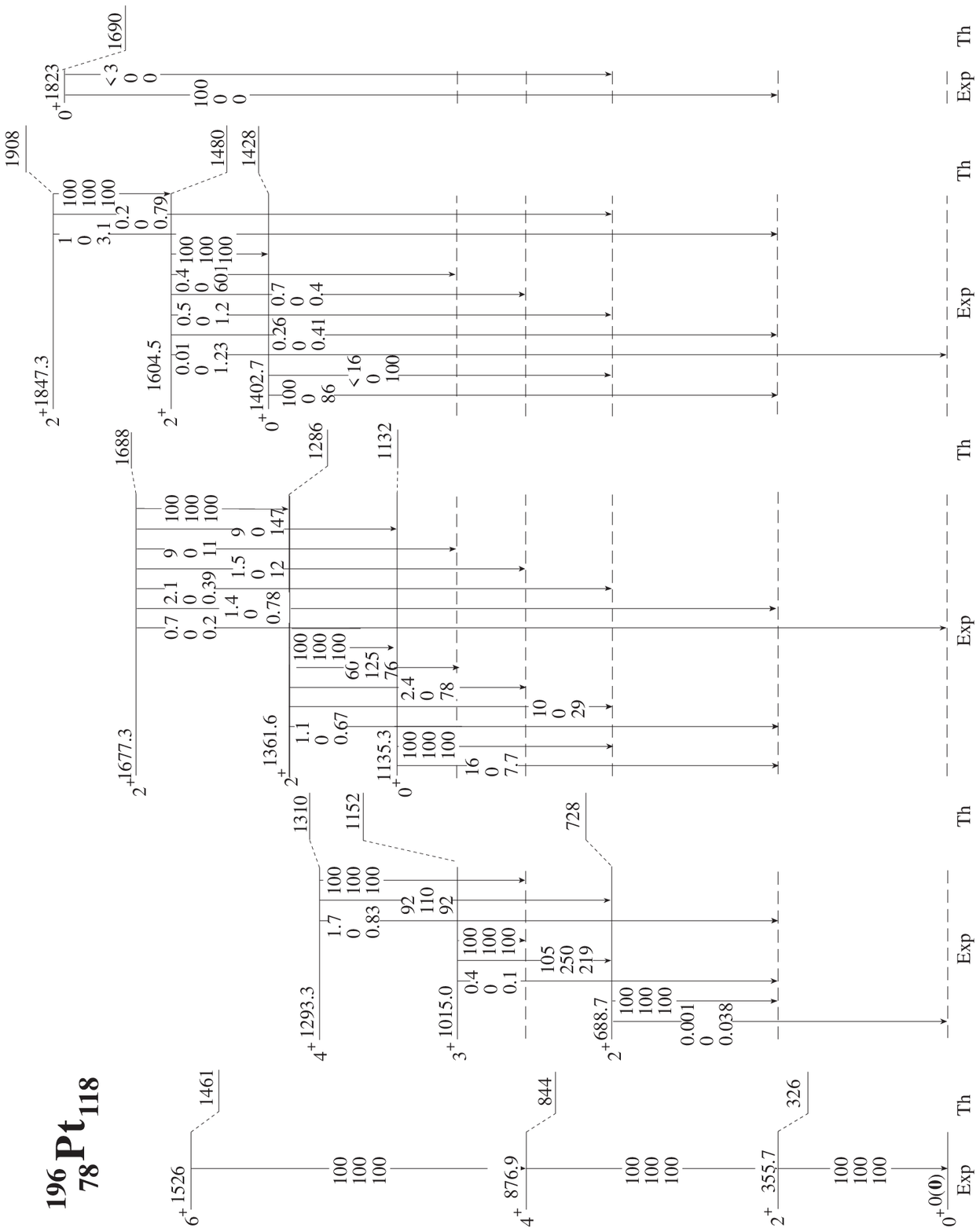}, we have presented a separate spectrum for
$^{196}$Pt, in order to show the experimental $E2$ branching
 ratios as well as the corresponding {\small FDSM} and {\small IBM--2}
predictions.  The overall {\small FDSM} results are in quantitative
agreement with the $O(6)$ level pattern and the \so6 $ E2$ selection
rules in the {\small FDSM}
 description of $^{196}$Pt. This agreement confirms that the
{\small FDSM}  does not have an explicit mathematical $SO(6)$ dynamical
symmetry for this case, but
 it has a remarkably accurate practical one \reference{fen93}.

\subsubheading{em-tran}{Electromagnetic Transitions}

The {\small FDSM} wavefunctions obtained through the diagonalization of
the Hamiltonian can be used to compute  multipole transition
strengths.  Expressions for quadrupole moments and $B(E2)$ values,
dipole moments and g-factors, and isomer and isotope shifts may be
found in Ref.\ \reference{pin95}.  In Table \ref{tb:pt196be2}, we have
compiled the experimental $B(E2)$ values and predictions from the
{\small FDSM} and various versions of the {\small IBM} for $^{196}$Pt.
Additional comparisons for the other isotopes of Pt may be found in
Ref.\ \reference{pin95}.  Also shown in the $^{196}$Pt table are \so6
limit results.  ({\small IBM--1(g)}  means {\small IBM--1} with one
g-boson explicitly included and the effective charge  individually
fitted from nucleus to nucleus;  in the {\small IBM--2} calculation
the effective charge remains  constant for all the platinums, just as
for the {\small FDSM} calculations.) Table \ref{tb:gfactors} compares
experimental g-factors with  {\small FDSM} and {\small IBM}
calculations, and calculated isomer and isotope shifts are compared
with data in \tableref{isomer}.

\subsubheading{fspin}{F-Spin and Majorana Interactions}

 In boson-model calculations,  $F$-spin is used to distinguish
low-lying symmetric and mixed symmetry states \reference{ots77}, and a
``Majorana interaction'' is typically introduced to guarantee that
mixed symmetry states do not appear at unacceptably low energies.  For
fermions there is no closed $F$-spin algebra and no explicit Majorana
interaction in the  Hamiltonian: for a fermion many-body system, the
relative positions of the symmetric and mixed-symmetry states are
governed by the basic pairing and multipole interaction strengths.  For
the $^{196}$Pt case that we discuss, the mixed-symmetry $1^{+}$ state
is at an excitation energy of approximately 2 MeV, and is correctly
predicted without introducing the analog of a Majorana term. Thus, the
effective fermion $SO(6)$ symmetry produces a quantitative agreement
with the spectrum using a simpler Hamiltonian than is required in the
{\small IBM--2}.

\subheading{5.3.2}{Effective Interaction for Rare Earth Nuclei}

The parameters used in reproducing the $SO(6)$-like spectrum and the
branching ratios of the platinum isotopes do not represent a peculiar
set chosen solely for this purpose. Similar calculations in the
remainder of the rare earth nuclei indicate that the Pt parameters are
related to the effective interactions of the entire shell. For example,
if the same five parameters are used in the {\small FDU0}
 calculation but the appropriate valence particle numbers for heavy Gd
isotopes are chosen, the spectrum and transition rates become
$SU(3)$-like (axially-symmetric rotors).  With only small adjustments
in these parameters (for example, less than 5\% for the n--p quadrupole
coupling, which is the dominant parameter) the spectrum and transition
rates for the low-lying states in the heavier Gd isotopes can be {\em
quantitatively} reproduced. Likewise, a similar set of parameters leads
to vibrational ($SU(2)$-like) spectra for valence particle numbers near
the beginning or the end of the shell.

These results imply that a fixed Hamiltonian with parameters varying
slowly with particle number can produce spectra that evolve from
$SU(2)$ to $SU(3)$, from $SU(3)$ to $SO(6)$, and from $SO(6)$ to
$SU(2)$, as the proton or neutron number changes. This
$SU(2)$--$SO(6)$--$SU(3)$ triangle is similar to the
$SU(5)$--$O(6)$--$SU(3)$ triangle relation \reference{cas88} of the
{\small IBM}.
 However, unlike the {\small IBM,} the triangle relationships in the
{\small FDSM}  are {\em shell dependent}. For the $SO(8) \times SO(8)$
shells, there is no $SU(3)$ limit and the $SO(6)$ symmetry is the most
collective limit (corresponding to $\gamma$-unstable rotations). In
this case the triangle relationship is replaced by
$SU(2)$--$SO(7)$--$SO(6)$.  We have speculated \reference{fen93} that
whenever there is a vibrational ($SU(2)-$like) symmetry limit on one
corner of the triangle and an $SU(3)-$like rotational limit on a second
corner of the triangle, an $SO(6)$-like effective transitional symmetry
may occur, even though there is formally no such dynamical symmetry in
these shells.

\subheading{5.3.3}{Universality of Normal and Exotic States}

The  measured electromagnetic transition strengths $B(E2)$ between the
ground states and first $2^+$ states of even--even rare earth nuclei vary with
mass number $A$ in a way quite similar to that of  the summed orbital
$B(M1)$ strengths measured in the same nuclei \reference{ran91}.  The
observed correlation in these strengths suggests a microscopic
relationship between modes that are qualitatively different in the
simplest geometrical picture.  Assuming that all $M1$ strength has been
detected experimentally, it is difficult to explain simply within the
{\small IBM} $F$-spin symmetry the observation that both the $E2$ and
$M1$ strengths saturate well before midshell  \reference{gin91}.  The
{\small FDSM} is known to reduce to an {\small IBM} model in the limit
of neglected Pauli correlations \reference{wu94b}, so it is reasonable
to ask whether a model such as the {\small FDSM}
 that incorporates the fermionic nature of the correlated pairs
can account for the similar behavior of $E2$ and $M1$
 strengths.

As noted in the preceding section, the $F$-spin algebra of {\small
IBM--2} allows the introduction of an $F$-spin scalar (the Majorana
interaction) that is insensitive to the normal symmetric states.  Thus, It
may be chosen to have a strength that  pushes the mixed-symmetry states to
their proper energies.  Unlike the {\small IBM--2}, the {\small FDSM}
cannot have a closed $F$-spin algebra.    Thus, one is compelled to
describe all states, symmetric or mixed-symmetry, in terms  of a {\em single
Hamiltonian} whose eigenstates must reflect the subtle balance between
n--p quadrupole interactions  and the n--n and p--p pairing forces.

\subsubheading{m1e2ham}{Hamiltonian and Operators}

We employ a 5-parameter (pairing and QQ) FDSM Hamiltonian:
\begin{equation}
 H=G'_{0\pi}S^{\dag}_{\pi}S_{\pi}^{\phantom{\dag}}+
    G'_{0\nu}S^{\dag}_{\nu}S_{\nu}^{\phantom{\dag}}+
    B'_{2\pi} P^{2}_{\pi} \cdot P^{2}_{\pi}+
    B'_{2\nu} P^{2}_{\nu} \cdot P^{2}_{\nu}+
    B_{2\pi\nu} P^{2}_{\pi} \cdot P^{2}_{\nu} .
\tag{ham5p}
\end{equation}
The operators are defined in \reference{wu94b} and the {\small FDSM}
quadrupole pairing interaction is  taken into account by renormalizing
the parameters: $ G'_{0\sigma}=G_{0\sigma}-G_{2\sigma}$ and $
B'_{r\sigma}=B_{r\sigma}-G_{2\sigma}$ ($\sigma=\pi,\nu$).  The five
parameters in  \eq{ham5p} were determined numerically using a gradient
search with the {\small FDU0} code \reference{hwu89,val91} fitted to the
experimental energies of the available $2^{+}_{1}$, $4^{+}_{1}$,
$6^{+}_{1}$, $1^{+}$, $2^{+}_{2}$, and $0^{+}_{2}$ states in Nd, Sm,
Gd, Dy, and Er.  These calculations indicate that the QQ-interaction
plays the crucial role in correlating the energies of the $2^{+}_{1}$
and $1^{+}$ states.  Using these parameters, we may then ask whether
there is a correlation between symmetric and mixed-symmetry states caused by
the
same $Q_{\pi} \cdot Q_{\nu}$ strength for the electromagnetic
transitions.

\subsubheading{spectmul}{Spectrum and Multipole Strengths}

In \fig{E21}  the experimental and calculated energies of the
$2^{+}_{1}$ and $1^{+}_{1}$ state energies are plotted versus the $P$
factor \reference{cas87,ran91}, $P\equiv N_{p}N_{n}/(N_{p}+N_{n})$,
where $N_{p}$ ($N_{n}$) are the valence proton (neutron) numbers,
respectively.  The $B(E2)$ values and the summed $B(M1)$ strengths
resulting from this calculation are compared with measured quantities
in \fig{e2m1}.  The curves are the empirical relations presented in
\reference{ran91} that summarize the approximate behavior of the data.
The $B(E2)$ and $B(M1)$ strengths are reproduced {\em quantitively}
 by the calculations (with the single exception of the $^{164}$Dy
point).  Thus, we find theoretical evidence for the approximate
universal behavior of $E2$ and $M1$ strengths exhibited by the data.
Furthermore, even the deviations from universality exhibited by the
data are reproduced by the calculations without parameter adjustment.
In Figs.\ \ref{fg:e2m1}c and \ref{fg:e2m1}c$^\prime$, we have plotted
the ratio $E(4^{+}_{1})/E(2^{+}_{1})$ as a function of $P$.  This
quantity also varies with $P$ in a way similar to that of the $E2$ and
$M1$ strengths, and  is also quantitively reproduced by these
calculations.

Finally, we address the question of how important the variation of
effective interaction parameters is to the success of the calculations
we have described.  In
Figs.\ \ref{fg:e2m1}a$^{\prime\prime}$--c$^{\prime\prime}$ we repeat
the calculations of Figs.\ \ref{fg:e2m1}a$^{\prime}$--c$^{\prime}$, but
with a {\em fixed set of parameters for all nuclei:}
$G'_{0\pi}=-0.074$  MeV, $G'_{0\nu}=0.020$ MeV, $B'_{2\pi}=-0.001$ MeV,
$B'_{2\nu}=0.047$  MeV, and $B_{2\pi\nu}=-0.243$ MeV.  The results of
this calculation are seen to differ  in minor details only from the
previous calculations in which the effective interaction was permitted
to have a weak $A$ dependence.  Thus, the quantitative reproduction of
$E2$ and $M1$ strengths in the rare earth nuclei is an inherent feature
of the FDSM with a constant effective interaction.

\heading{6}{Higher Heritage Configurations}

The {\small FDSM} is a truncated shell model with
symmetry-dictated $S$ and $D$ pairs. In its simplest implementation,
one restricts the space to
heritage $u=0$; then all nucleons form $S$ or $D$ fermion pairs:
\begin{equation}
  A^{\dagger r}_{\mu} = \left\{ \begin{array}{ll}
  \sqrt{\Omega_{1}/2}
     \left[ b^{\dagger}_{ki} b^{\dagger}_{ki} \right]
                 ^{0r}_{0\mu} ,~~~r=0,\, 2  &~ \mbox{for i-active} \\
   &  \\
   \sqrt{\Omega_{1}/2}
     \left[ b^{\dagger}_{ki} b^{\dagger}_{ki} \right]
                 ^{r0}_{\mu 0} ,~~~r=0,\, 2  &~ \mbox{for k-active}
  \end{array} \right.  ~~,
\end{equation}
where $i=\frac{3}{2}$ or $k=1$,
$S^{\dagger}=A^{\dagger 0}$ and $D^{\dagger}_{\mu}=A^{\dagger 2}_{\mu}$, and
$\Omega_{1}$ is
the degeneracy for the normal-parity levels of a major shell.

As we have seen, this space gives a good description for even--even
nuclei in states of low angular momentum.  However, numerical
calculations within the $ u=0$  space consistently overestimate the
energies and underestimate the $B(E2)$ values for higher angular
momentum states, and these discrepancies increase with angular momentum.
Some improvement can be gained in the energies by correcting
perturbatively for the influence of pairing in the same manner as the
$SO(5)$ + pairing approach discussed previously.  However, it has been
realized since the inception of the {\small FDSM} that the primary
reason for these discrepancies is the significant contribution of
broken pairs to high angular momentum states \reference{gui86,gui87}.
Broken pairs in normal or abnormal parity orbitals may carry
significant angular momentum, thereby reducing the amount carried by
the $S$--$D$ condensate.  As was demonstrated schematically in Refs.\
\reference {gui86,gui87}, this leads to a Variable Moment of Inertia
({\small VMI}) behavior for the moment of inertia similar to that
observed experimentally, and brings calculated and observed high-spin
$B(E2)$ values into quantitative agreement with data
\reference{gui87,wu90c,wu94b}.

Thus, it is natural to extend the numerical implementation of symmetry
dictated truncation by using the {\small FDSM} to include unpaired
particles.  This has been discussed for a single unpaired particle in
Refs.\ \reference{hwu87,hwu87b,hwu88,wu94b}. We shall not discuss that
further here, but instead conclude by summarizing recent work
that incorporates broken pairs in such calculations.  We have developed
a computer code {\small SU3}su{\small 2} that includes explicit broken pairs in
an $SU(3)$ limit in order to examine the yrast states in the rare earth
and actinide regions \reference{di96,pan96}.  In this extension, the
model basis is constructed by coupling the $SU(3)$ basis and one
broken pair in the normal parity levels or the
unique parity level
\begin{equation}
  (S,D)^{N-1}\otimes A'(i) ~~~ \mbox{or}~~~ (S,D)^{N-1}\otimes A'(j_{0}).
\end{equation}
where $S$ and $D$ are the usual symmetry-dictated
coherent fermion pairs with
coupled angular momentum 0 and 2, respectively,
$A'(i)$ designates
broken normal-parity pairs,
and $A'(j_{0})$ designates broken pairs
in the unique parity level.
The corresponding creation operators are
\begin{equation}
A'^{\dagger}(i)= \sum _{ki}\sqrt{\Omega_{1} /2} \left[
  b_{ki}^{\dagger}b_{ki}^{\dagger}
  \right] ^{(KI)L'}_{(M_{K}M_{I})M_{L'}},
\end{equation}
\begin{equation}
A'^{\dagger}(j_{0})= \sum _{ki}\sqrt{\Omega_{1} /2} \left[
  b_{j_{0}}^{\dagger}b_{j_{0}}^{\dagger}
  \right] ^{I_{0}}_{M_{I_{0}}}.
\end{equation}
The basis for the $SU(3)$ core is
\begin{equation}
\begin{array}{ccccc}
Sp(6)& \supset&  SU(3) &\supset &SO(3)
\\[0.2em]
N_{1},u=0 & &(\lambda ,\mu)&&
		    \kappa L     \nonumber
\label{eq:su3}
\end{array}
\end{equation}
where $\kappa$ is an additional quantum number to
distinguish  orthogonal
states  having the same $(\lambda , \mu)$  and $J$.
For an $SU(3)$ core  and one broken pair, the
following coupling schemes are possible:
 $$
|N_{1}u=0(\lambda_{1},\mu_{1})\otimes (2,0);
 (\lambda_{2},\mu_{2})KI JM_{J} \rangle ,
$$
\begin{equation}
 |N_{1}u=0(\lambda_{1},\mu_{1})\kappa L\otimes K_{i};KI JM_{J} \rangle,
\end{equation}
 $$
|N_{1}u=0(\lambda_{1},\mu_{1})\kappa L \otimes I_{0};JM_{J} \rangle.
$$
and the total Hamiltonian is
\begin{equation}
  H = H\tsub{SD} + H_{A'} + H\tsub{mix},
\end{equation}
where $H\tsub{SD}$ is the $SU(3)$-plus-pairing Hamiltonian, $H_{A'}$
corresponds to the broken pair, and $H\tsub{mix}$ is the interaction between
the $(S,D)$ core and the broken pair.  The last term $H\tsub{mix}$
leads to  the mixing of heritage. Explicitly, the Hamiltonian can be
expressed as
\begin{eqnarray}
 H \align= \alpha K \cdot K + \gamma
  I \cdot I - \delta I\cdot K
  +\gamma_{0} I_{0}\cdot I_{0}
  -\delta_{0} I_{0}\cdot K                  \nonumber
\\
\align- \kappa P^{2}(k)\cdot P^{2}(i)
  +\kappa ' (D^{\dagger} D)^{2}\cdot P^{2}(i)
  -\kappa_{0} P^{2}(k)\cdot P^{2}(i_{0})
  + \kappa '_{0} (D^{\dagger} D)^{2}\cdot P^{2}(i_{0})   \nonumber
\\
\align+ \Delta B  ~~,
\label{eq:H}
\end{eqnarray}
where $\Delta B$ is the energy required to break one pair
in either normal parity  or abnormal parity levels.

In \fig{er1}, we present the results of a $u=2$ calculation of
the energy levels up to $J=20$ for the even--even isotopes
$^{160-166}$Er \reference{pan96}.  The inclusion of a single broken
pair in the \su3 symmetry limit leads to a quantitative description of
the spectrum.  We do not show results for $B(E2)$ values from these
calculations because that work is still in progress.  Preliminary
indicatons are that the inclusion of a single broken pair leads to a
substantial increase in the $B(E2)$ strengths at higher spins, but the
calculated values are still low in the angular momentum 10--20 region,
suggesting that configurations having two broken pairs may be important
for a quantitative description of $B(E2)$ values near angular momentum
20 for rare earth nuclei.

\heading{7}{Conclusions}

The Fermion Dynamical Symmetry Model ({\small FDSM}) provides a
systematic method for truncating the spherical shell model.  In this
scheme, a valence space is  selected using energy considerations and
principles of dynamical symmetry are then used to radically truncate
the valence space.  We term this  a symmetry-dictated truncation.  The
resulting  truncated space  permits systematic shell model calculations
to be implemented for all heavy nuclei.  Since the space  has been
severely truncated,  the corresponding interactions  are highly
effective with respect to the original shell model. Thus, the first
step in  systematic calculations with this truncation scheme is to
determine the appropriate {\small FDSM} effective interaction for each
valence space.  Although it is of considerable interest in the longer
term to relate such effective interactions to standard shell model
ones, the most practical initial way to determine the required
interaction is to construct it phenomenologically by viewing its matrix
elements as parameters to be constrained by a carefully chosen data set.  We
have presented examples of systematic {\small FDSM} calculations  that
have been used to determine an effective interaction appropriate for
configurations in heavy nuclei with no broken pairs.  This interaction
appears to be simple and to have a rather weak dependence on particle
number within major-shell valence spaces.  Calculations using this
interaction reproduce low-lying spectra, moments, and transition rates
for broad ranges of nuclei that exhibit varied collective behavior:
axial rotors, anharmonic vibrators, and gamma-soft rotors.  Finally, we have
presented an initial extension of this approach to include broken pairs in the
configuration space.  These
results constitute a practical demonstration that systematic shell
model calculations  are now feasible for  very heavy nuclei far removed
from closed shells.  The primary remaining task is to systematize and
refine the interaction over all mass numbers through a detailed series
of numerical calculations.

\noindent

\vspace {10pt}
\noindent
\paragraph{Acknowledgements.}
\drexelblurb
\taiwanblurb
\utkblurb
\ornlblurb

\newpage
\runningheads{\lefthead}{\righthead}

\bibliography{sd}          

\clearpage

\epsfiggy
{trun1}
{5.93in}
{10pt}
{0pt}
{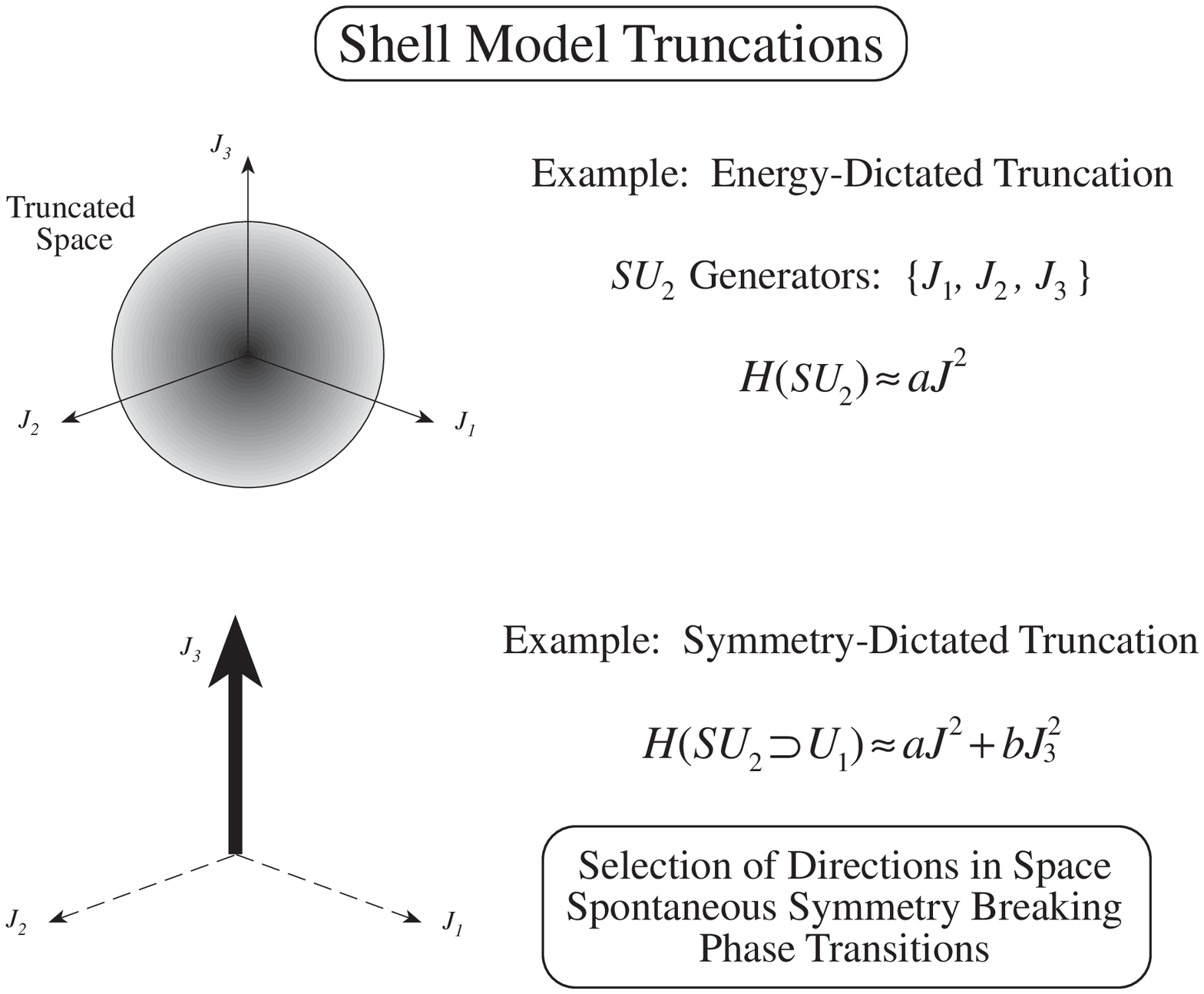}
{550}
{2.2in}
{\protect\baselineskip=20pt
An illustration of energy-dictated and symmetry-dictated truncations for a
simple symmetry.}

\epsfiggy
{trun2}
{5.29in}
{10pt}
{0pt}
{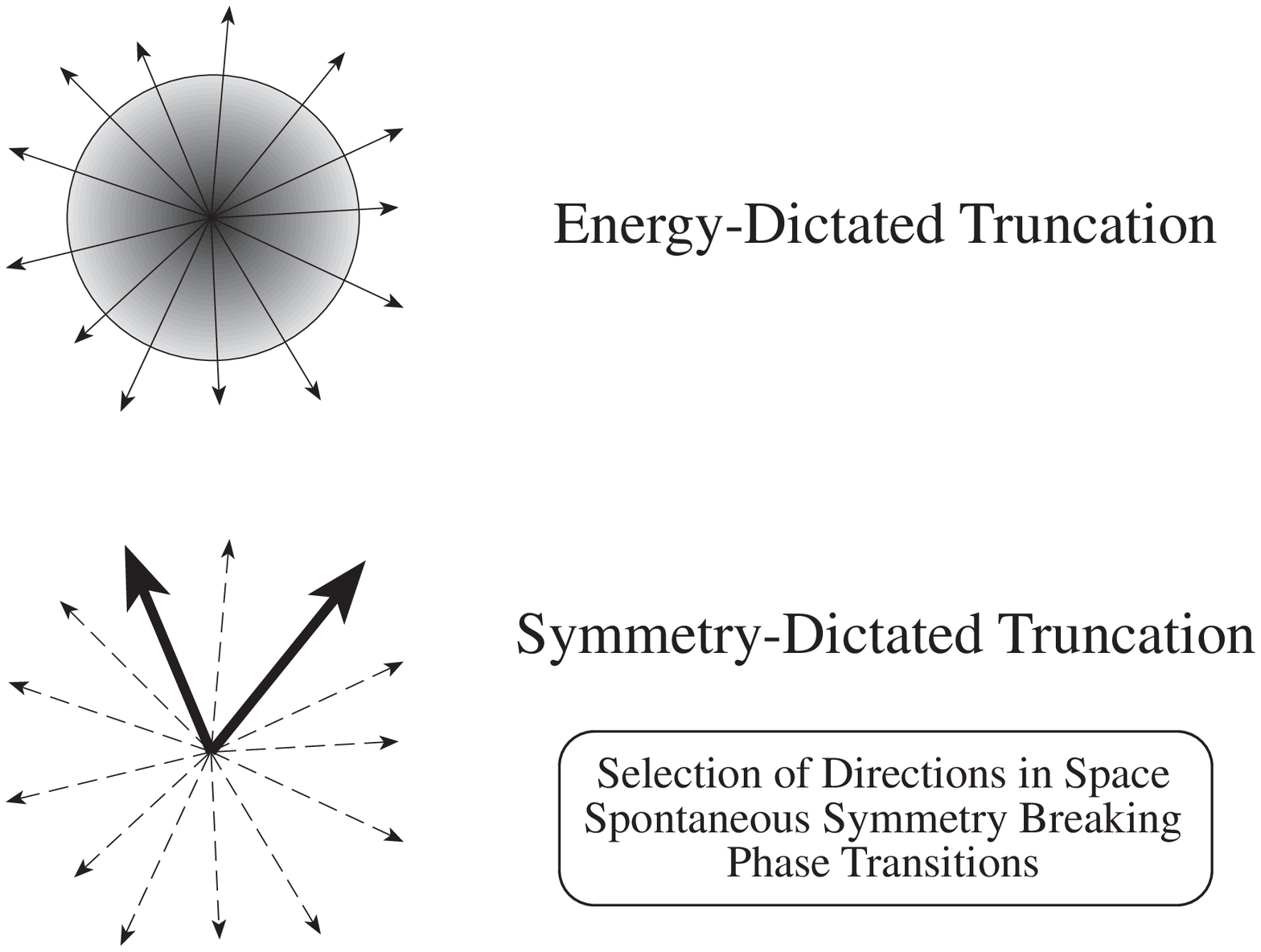}
{500}
{2.4in}
{\protect\baselineskip=20pt
Energy-dictated and Symmetry-dictated
truncations for a spherical shell
model.}

\epsfiggy
{belyaev5}
{4.28in}
{10pt}
{0pt}
{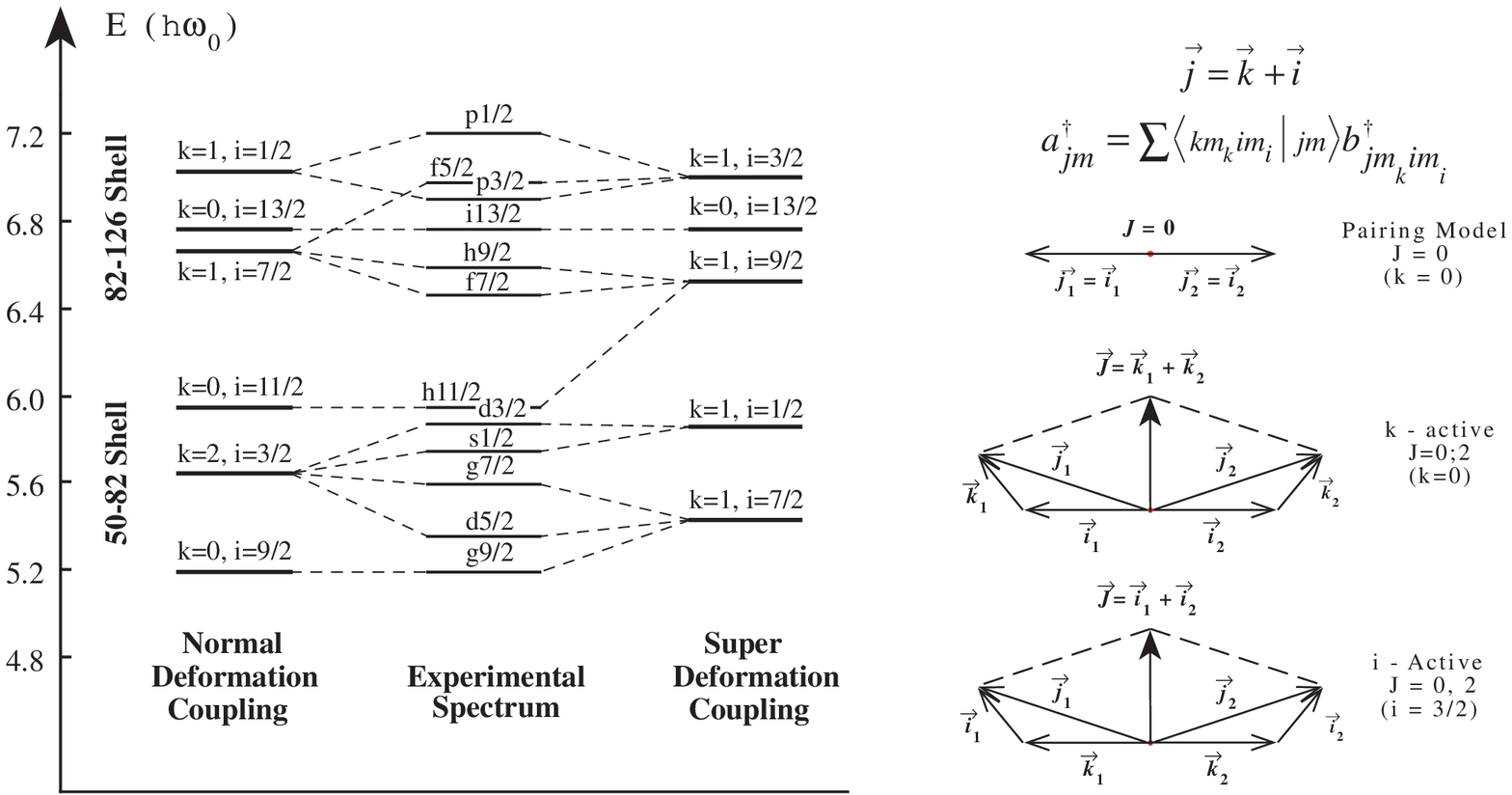}
{625}
{1.5in}
{\protect\baselineskip=20pt
The {\protect\small FDSM} coupling scheme for normal and superdeformation.}

\epsfiggy
{belyaev8}
{4.64in}
{10pt}
{0pt}
{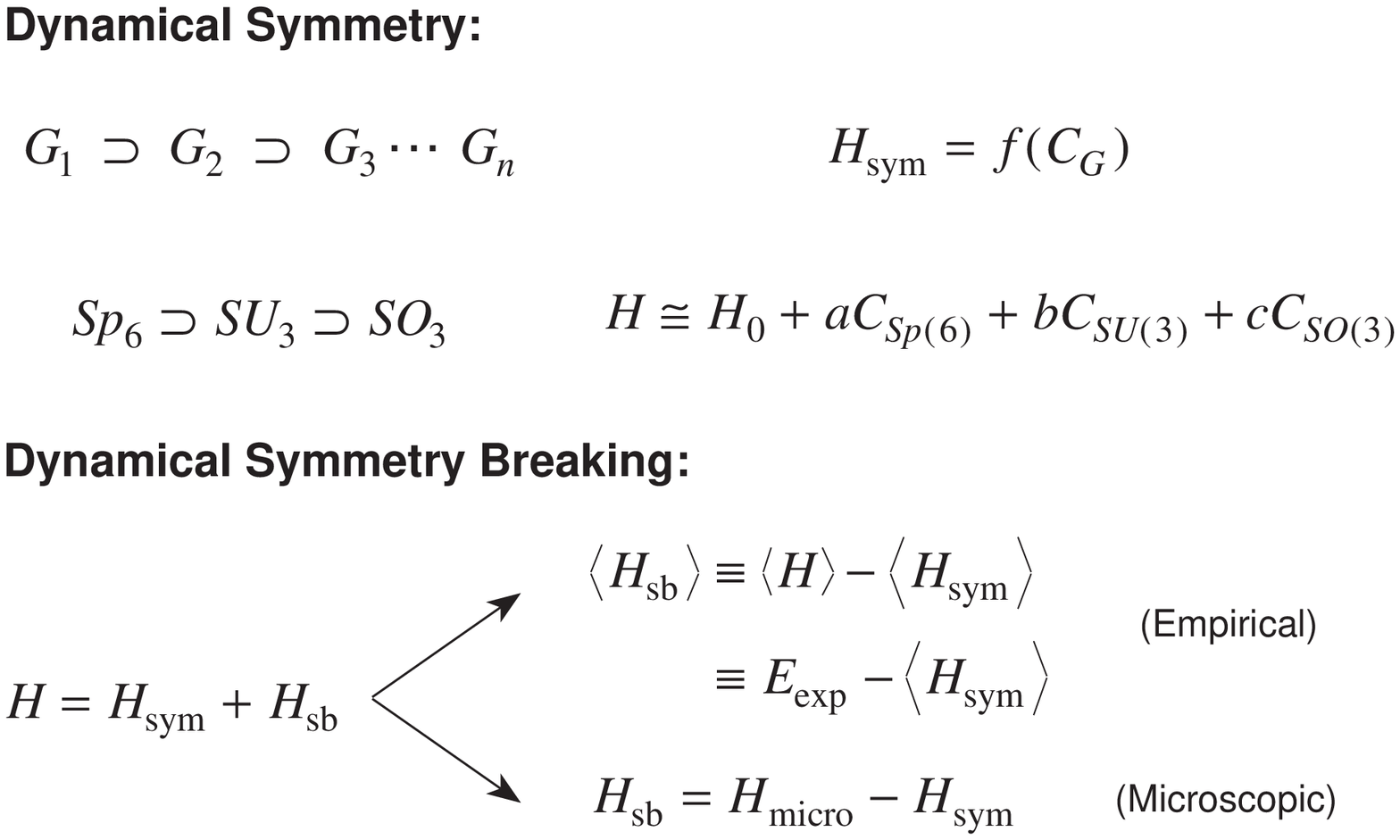}
{570}
{2.0in}
{\protect\baselineskip=20pt
Dynamical symmetry and the breaking of dynamical symmetry.  The Casimir
invariants of the group $G$ are denoted by $C_G$.}

\epsfiggy
{belyaev6}
{5.56in}
{10pt}
{0pt}
{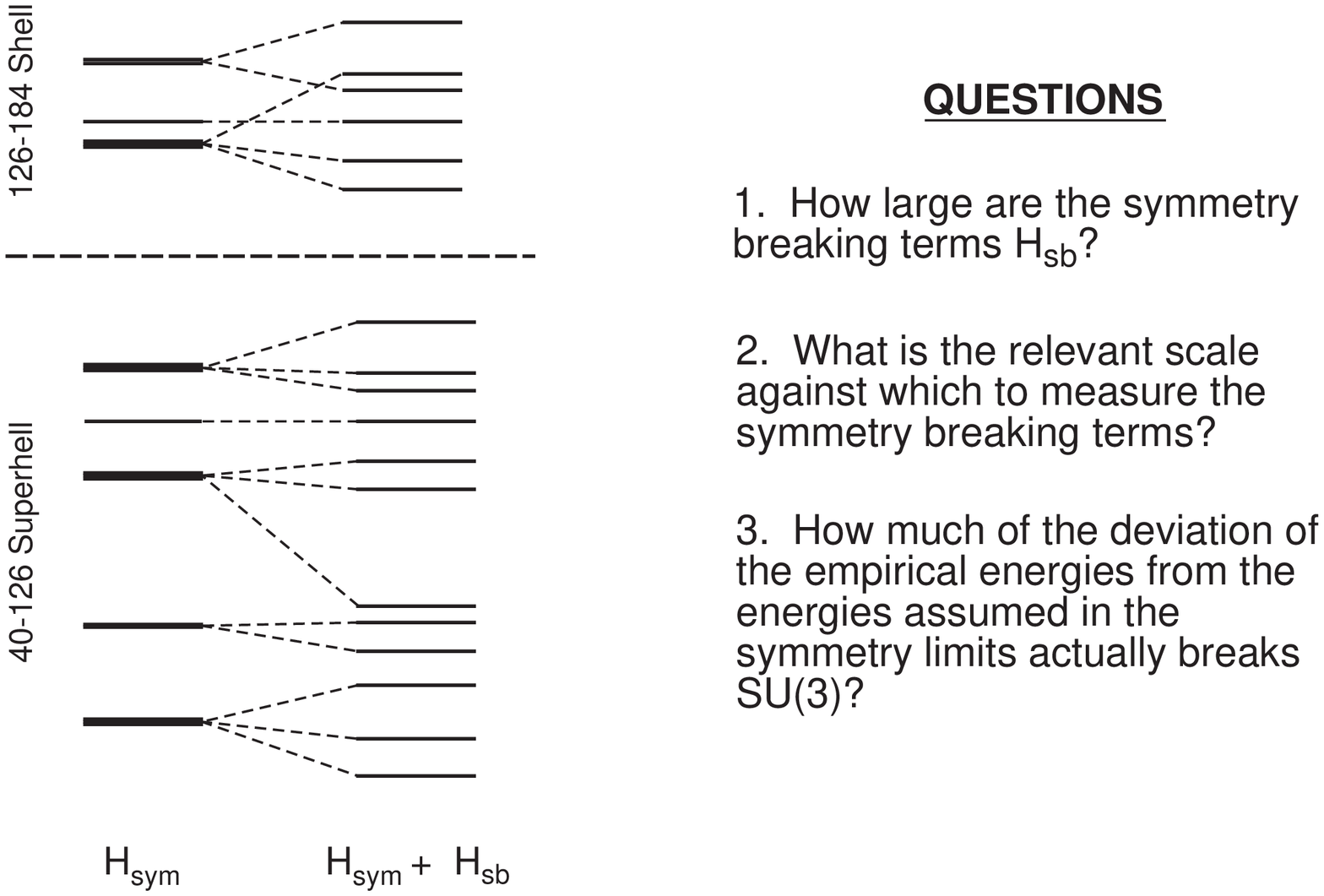}
{500}
{2.0in}
{\protect\baselineskip=20pt
The difference between the symmetry-limit and realistic single-particle
spectrum.}

\epsfiggy
{belyaev1}
{4.46in}
{10pt}
{0pt}
{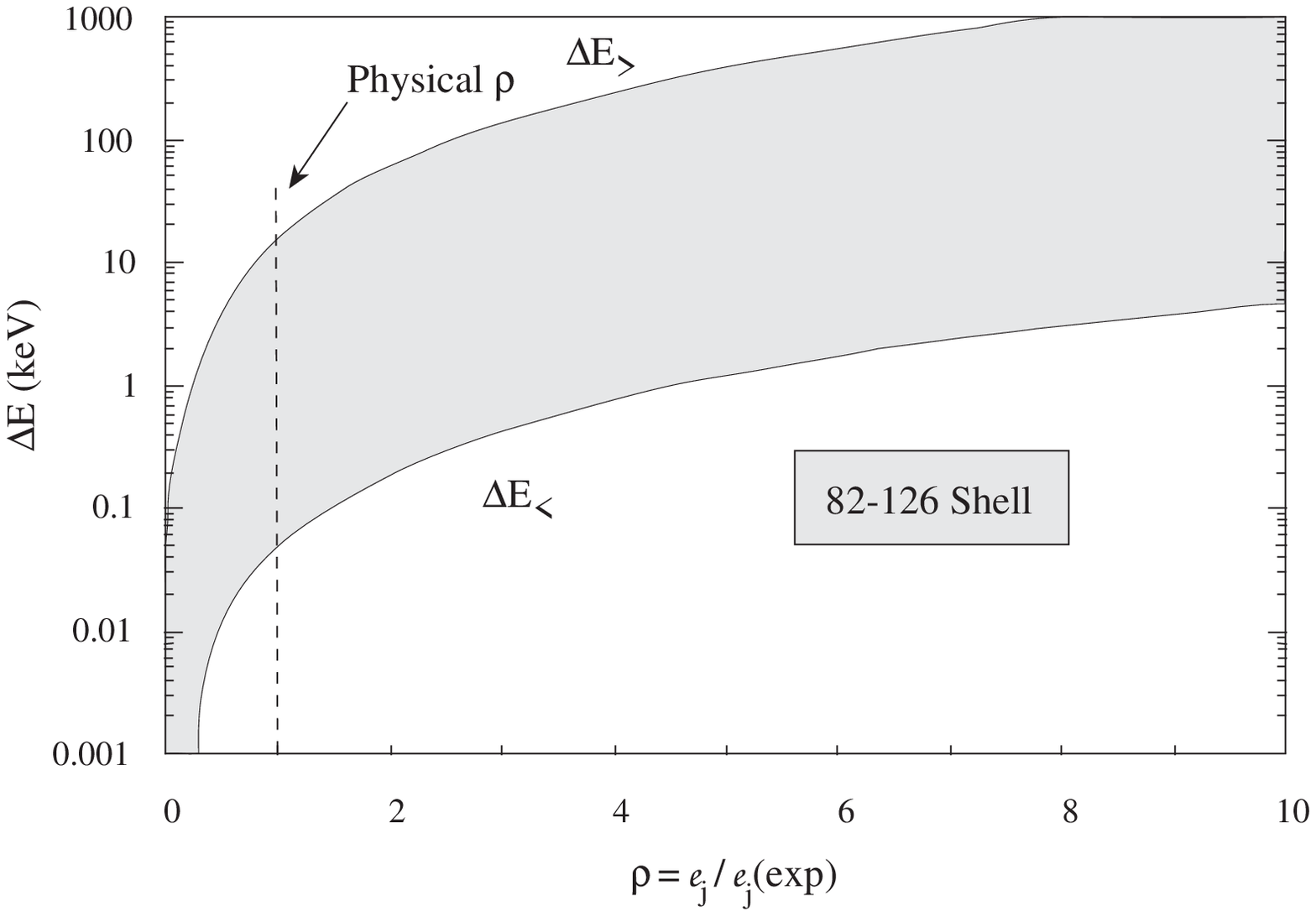}
{700}
{1.8in}
{\protect\baselineskip=20pt
Range of upper limits for $SU_3$
symmetry breaking by single-particle energies
in the 82--126 shell.}


\epsfiggy
{esurface}           
{4.35in}       
{10pt}             
{0pt}             
{shmin.eps}        
{825}            
{1.5in}
{\protect\baselineskip=20pt
Mass shell
correction for
heavy and superheavy elements using the dynamical symmetry methods of Ref.\
\protect\cite{han92}.}

\epsfiggy
{xeba1.eps}
{8.09in}
{10pt}
{0pt}
{xeba1.eps}
{960}
{2.3in}
{\protect\baselineskip=20pt
Comparison between calculated levels using eq.\ (\protect\ref{eq29a}),
and experimental energy levels for the even--even $^{120-126}$Xe isotopes.}

\epsfiggy
{xeba2.eps}
{6in}
{10pt}
{100pt}
{xeba2.eps}
{1100}
{2.2in}
{\protect\baselineskip=20pt
Comparison between calculated levels using eq.\ (\protect\ref{eq29a}) and
 experimental
 energy levels for the even--even $^{128-132}$Xe isotopes.}

\epsfiggy
{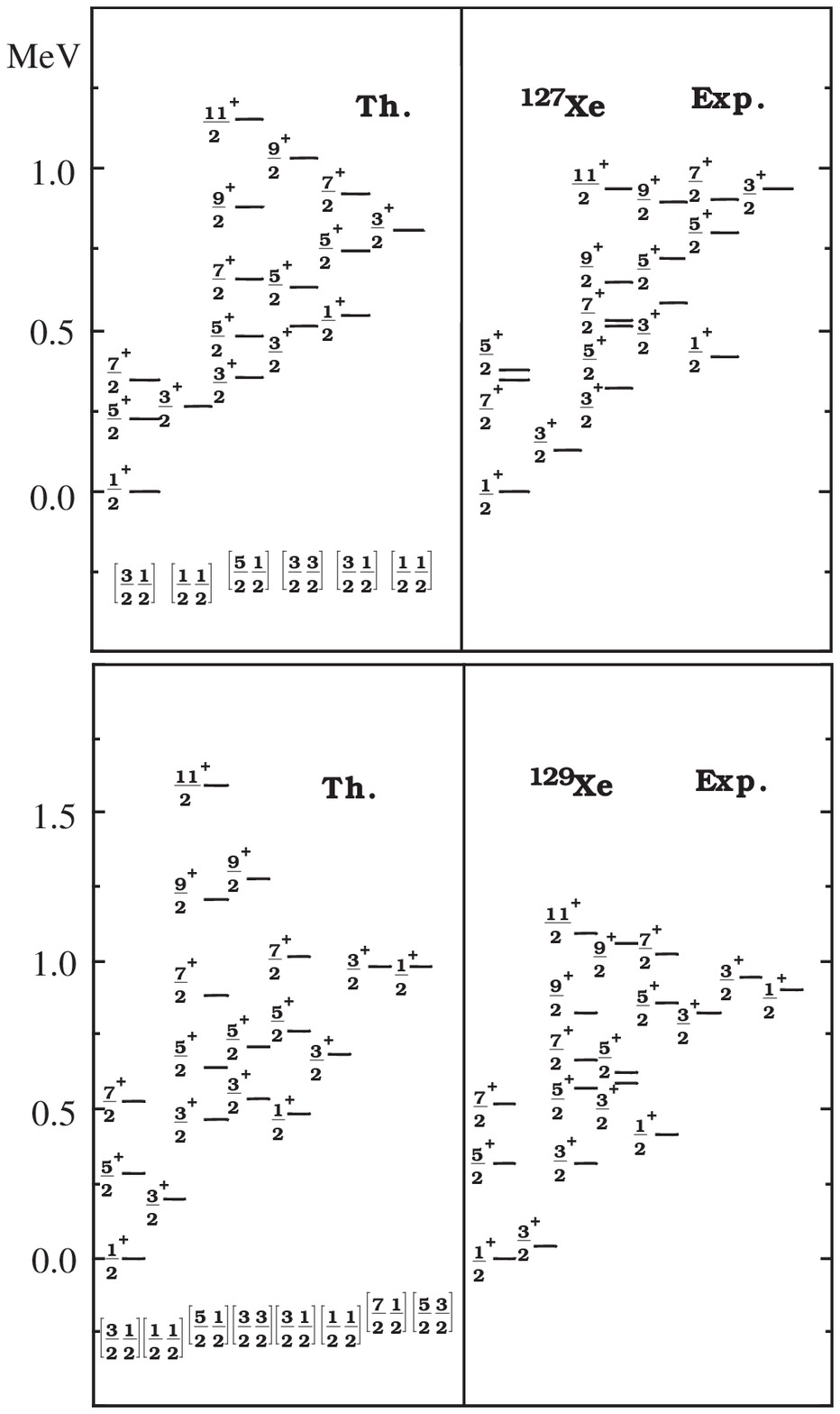}
{6.76in}
{10pt}
{0pt}
{xeba3.eps}
{1100}
{1.7in}
{\protect\baselineskip=20pt
Comparison between calculated levels using eq.\ (\protect\ref{eq29b}) and
 experimental energy levels for the even--odd $^{127-129}$ Xe isotopes. The
even--odd nuclei are constructed by coupling the neighboring even--even
core to a valence neutron.}

\epsfiggy
{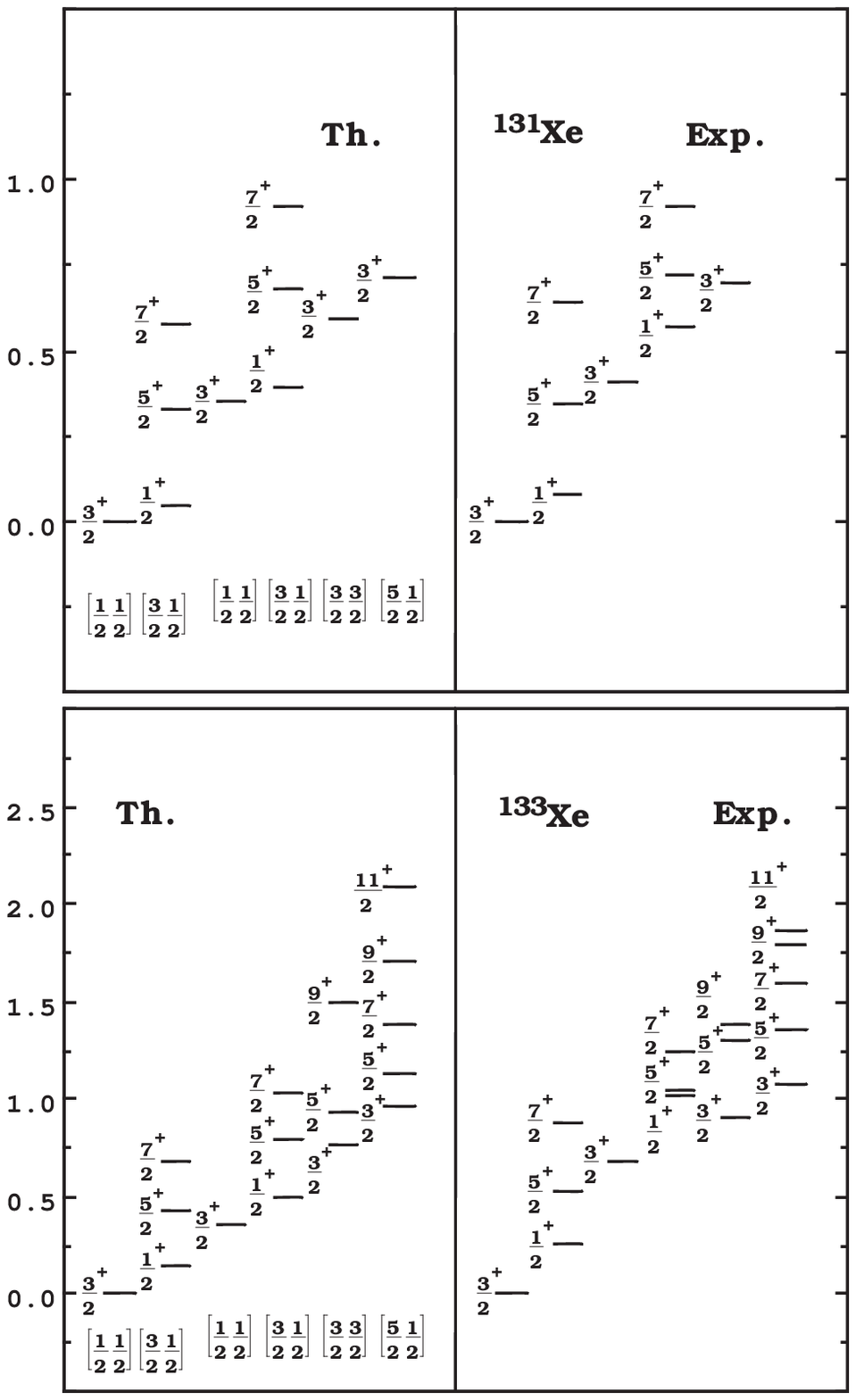}
{6.32in}
{10pt}
{100pt}
{xeba4.eps}
{1100}
{1.9in}
{\protect\baselineskip=20pt
Comparison between calculated levels using eq.\ (\protect\ref{eq29b}) and
 experimental energy levels for the even--odd $^{131-133}$Xe isotopes. The
even--odd nuclei are constructed by coupling the neighboring even--even
core to a valence neutron.}

\epsfiggy
{xeba6.eps}
{2.75in}
{10pt}
{0pt}
{xeba6.eps}
{1050}
{2.3in}
{\protect\baselineskip=20pt
The systematic shift of the ground band as a function of
 mass number for  Xe  isotopes.}

\epsfiggy
{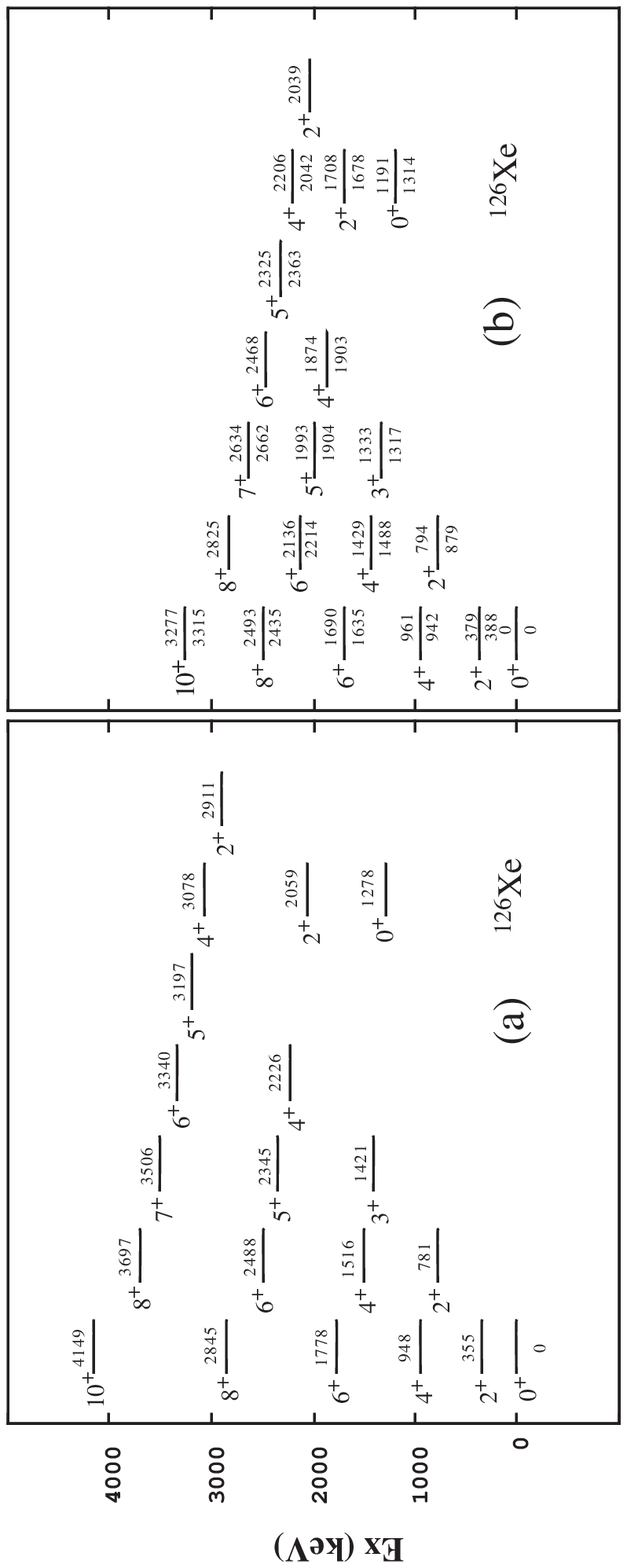}
{7.89in}
{10pt}
{0pt}
{xeba8.eps}
{1000}
{2.5in}
{\protect\baselineskip=20pt
The tau compression effect:  (a) calculated with  Eq.\ (\protect \ref{eq29a})
and (b) with Eq.\ (\protect\ref{eq216}).  Theoretical energy is shown above and
experimental energy below each level.}

\epsfiggy
{pt.fig2.eps}
{7.36in}
{10pt}
{10pt}
{pt.fig2.eps}
{1000}
{1.1in}
{\protect\baselineskip=20pt
Level scheme for positive-parity states in $^{196}$Pt. Experimental
levels  are taken from \protect\cite{ciz78}. }

\epsfiggy
{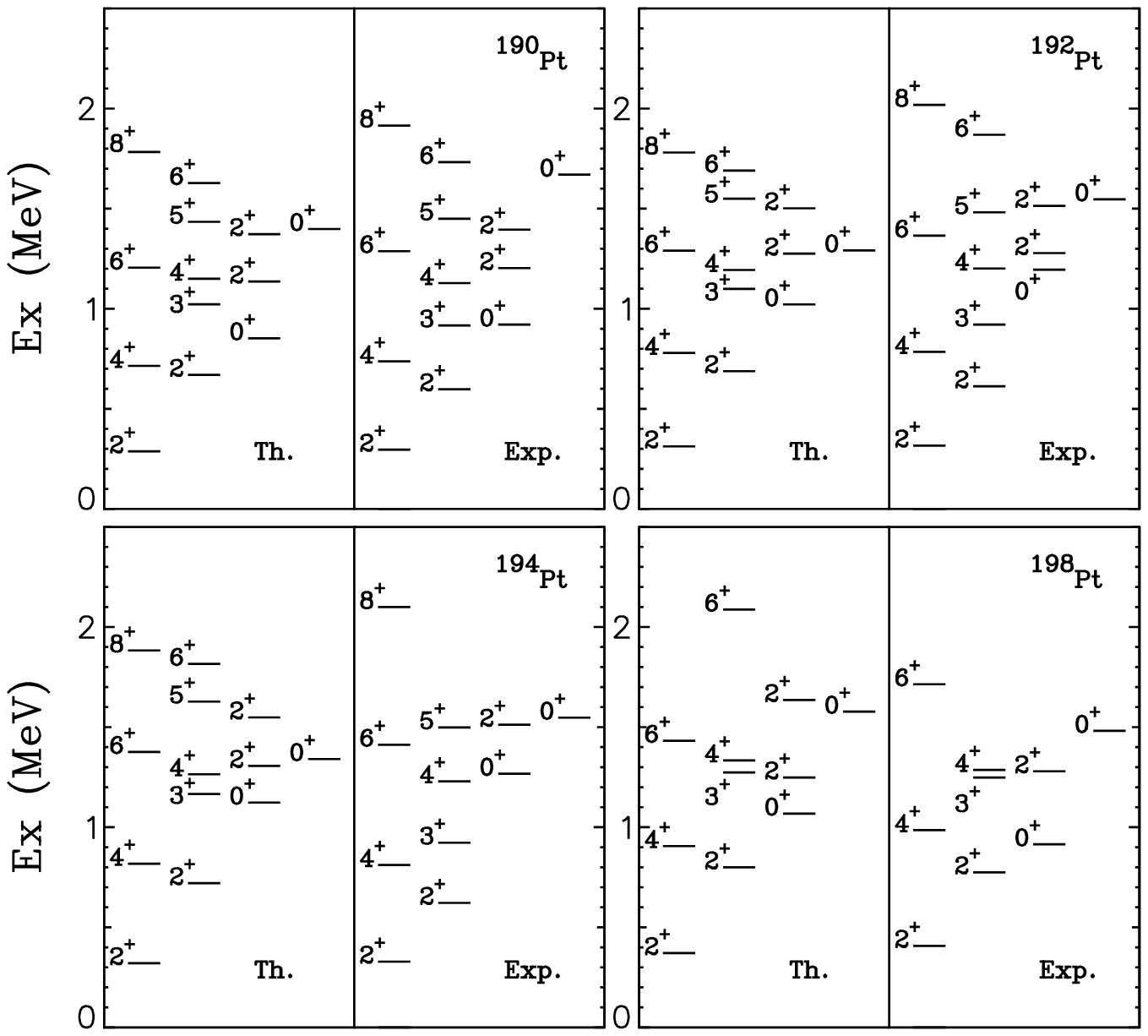}
{4.83in}
{10pt}
{0pt}
{pt.fig1.eps}
{800}
{1.9in}
{\protect\baselineskip=20pt
Spectra for positive-parity states in \protect\isotope{190-194,198}{Pt}.
Experimental levels are taken from
\protect\reference{sin89,sin90,zho90,shi91}.}

\epsfiggy
{E21}
{3.2in}
{10pt}
{0pt}
{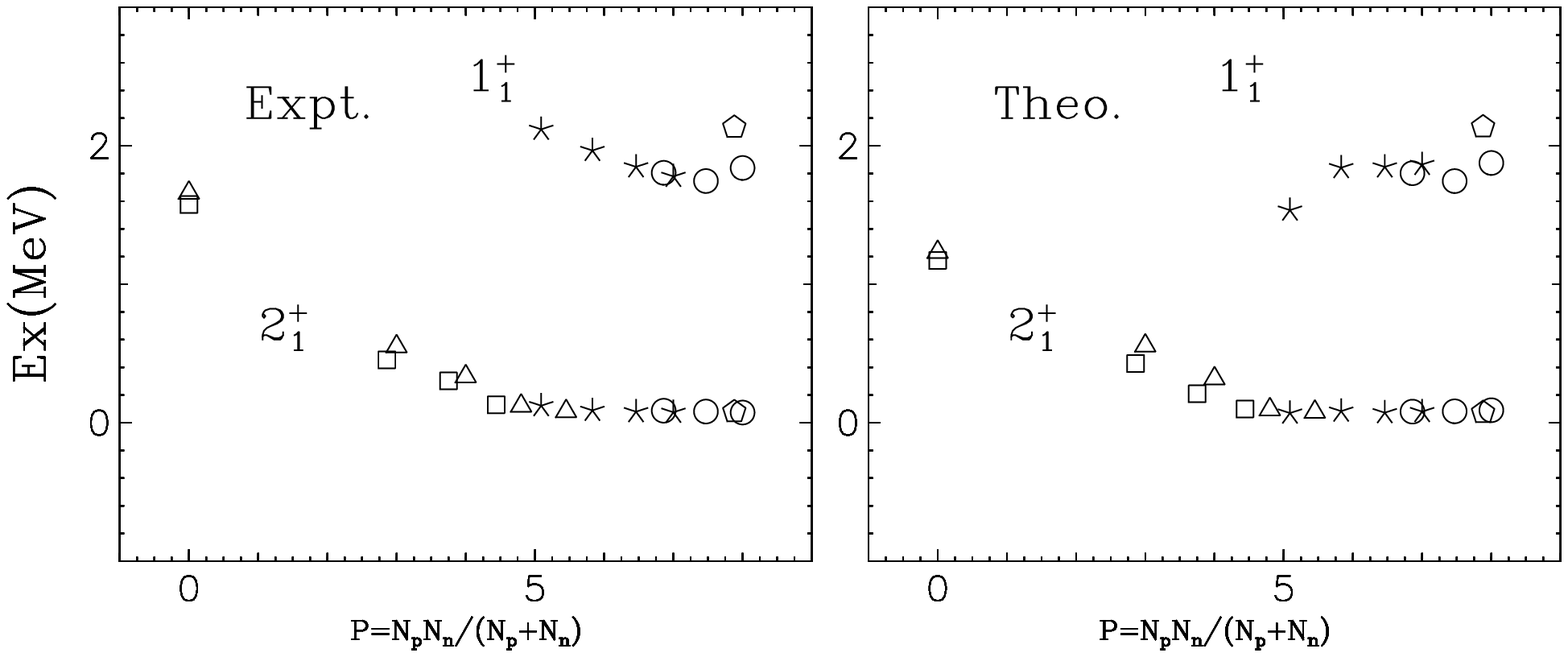}
{650}
{1.5in}
{\protect\baselineskip=20pt
Experimental and calculated energy levels for the first $2^+$ (lower
points) and $1^+$ (upper points) states in selected even--even rare
earth nuclei where orbital $M1$ strengths have been measured.  The
symbols denote isotopes with the same meanings as in
\protect\fig{e2m1}.}

\epsfiggy
{e2m1}
{5.92in}
{10pt}
{150pt}
{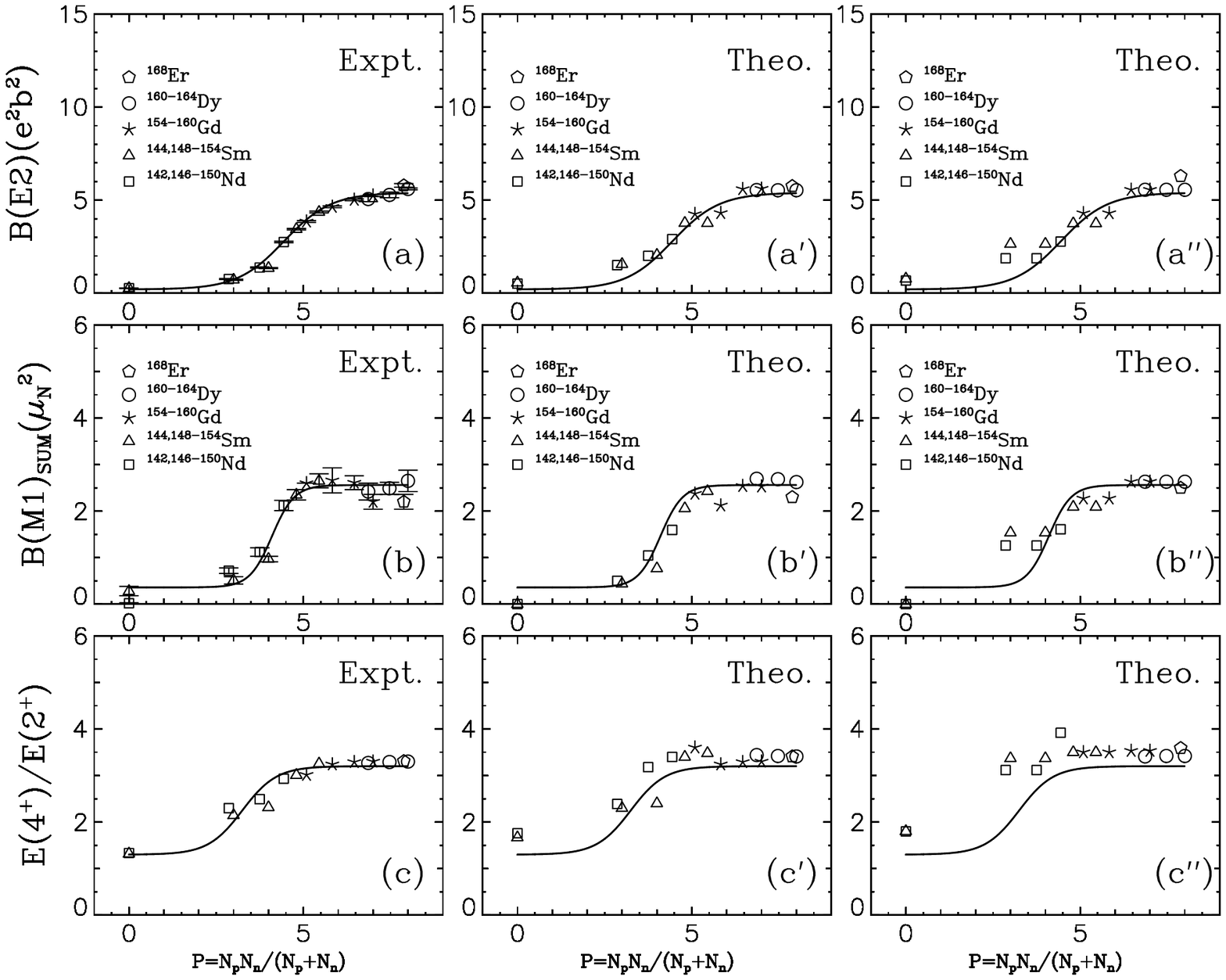}
{800}
{1.0in}
{\protect\baselineskip=20pt
Comparison of experimental  and theoretical $B(E2)$ and summed $B(M1)$
values for rare-earth nuclei as a function of $P$.  The experimental
results in the left column are taken from the compilation given in
\protect\reference{ran91} and \protect\reference{ram89}. For
comparison, curves from the same empirical relation employed there,
$B(E2,M1)=a_{1}+a_{2}/[1+exp((c-P)/d)]$ are also plotted. The symbols
in (c)--(c$^{\prime\prime}$) have the same meaning as in the $E2$ and
$M1$ cases.  The theoretical results in the second column of figures
correspond to best adjustment of effective interaction parameters to
reproduce spectra.  The theoretical results in the third column of
figures correspond to constant values of the effective interaction
parameters for all nuclei examined.}


\epsfiggy
{er1}           
{4.14in}       
{10pt}             
{0pt}             
{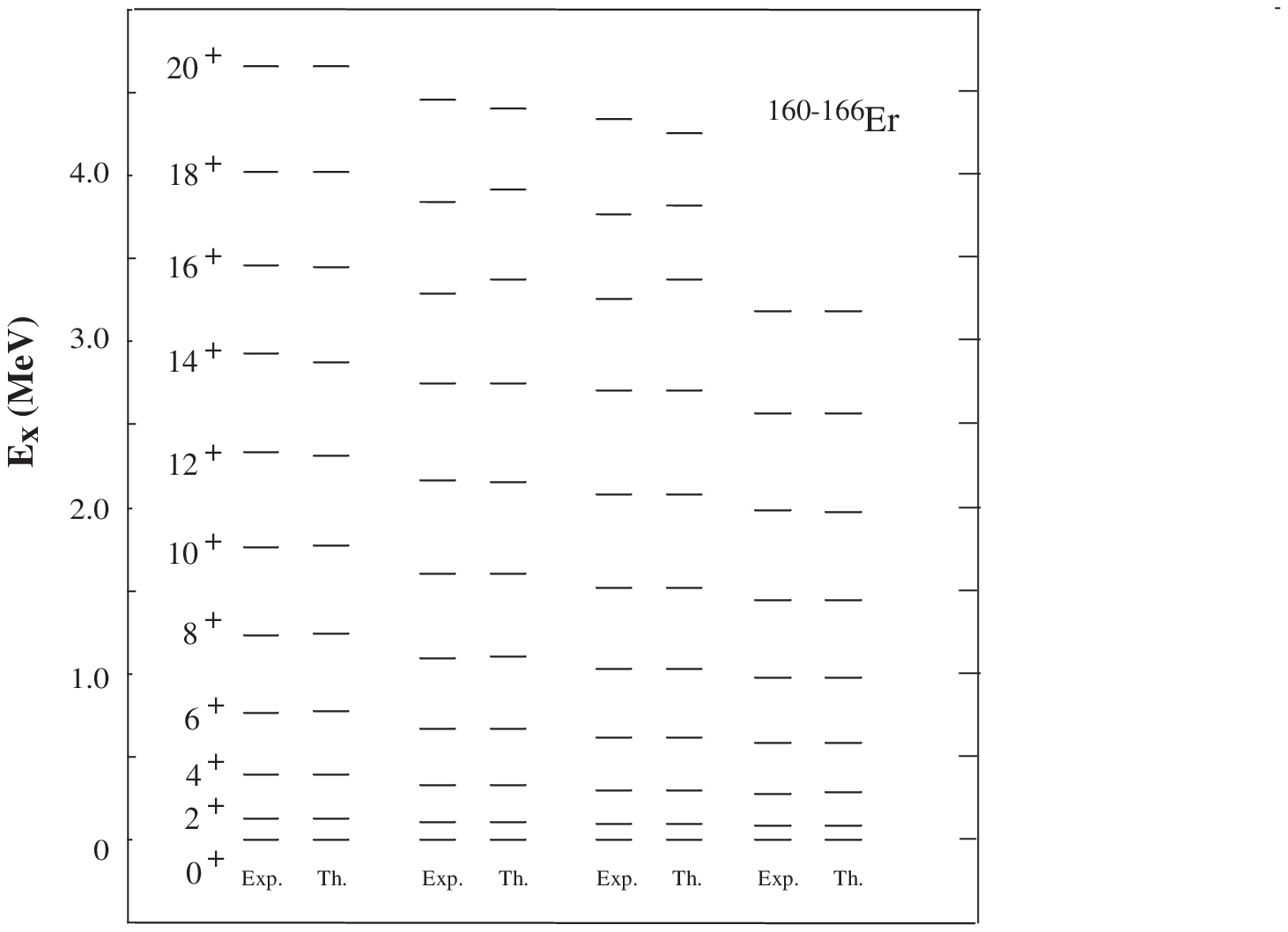}        
{1000}            
{1.7in}            
{\protect\baselineskip=20pt
Experimental and theoretical energy levels in
some even--even Er isotopes calculated with a single
broken pair using the code {\protect\small SU3su2}
\protect\reference{di96,pan96}.  The spectrum of
$^{160}$Er is to the left and that of $^{166}$Er to the right.}


\clearpage

\begin{table}[t]
\begin{scriptsize}
\begin{center}
\center\caption{\protect\small
Experimental and Theoretical Mass Excesses in
MeV for the Heaviest Elements.
\protect\label{tb:sheavy}}
\vspace{0.2em}
\begin{tabular}{lllllllll}
\heavyline
$^{256}104$&
$^{258}105$&
$^{260}106$&
$^{262}107$&
$^{264}108$&
$^{266}109$&
$^{269}110$&
$^{272}111$&
Reference
\\
\heavyline
94.25(03)&
101.84(34)&
106.60(04)&
114.68(38)&
119.82(30)&
128.39(35)&
134.80(32)&
141.70(37)&
Exp \cite{aud93,hof95}
\\
94.15&
101.98&
106.8&
114.82&
120.02&
128.16&
133.31&
140.49&
FDSM \cite{han92}
\\
95.90&
103.41&
106.87&
116.10&
121.28&
129.44&
&
144.83&
Myers \cite{ato76}
\\
94.84&
102.22&
105.81&
115.00&
120.40&
128.43&
&
144.04&
Groote \cite{ato76}
\\
95.6&
102.6&
106.8&
114.9&
120.4&
127.8&
&
142.6&
Seeger \cite{ato76}
\\
94.37&
101.64&
105.68&
114.78&
120.27&
128.44&
&
143.82&
Liran \cite{ato76}
\\
95.77&
103.01&
108.13&
115.50&
&
&
&
&
M\"{o}ller \cite{mol81b}
\\
95.69&
103.11&
108.12&
115.71&
121.09&
129.04&
135.51&
143.44&
M\"{o}ller \cite{mol88}
\\
93.78&
101.00&
105.81&
113.18&
118.34&
126.06&
132.39&
140.18&
M\"{o}ller \cite{mol88b}
\\
93.38&
100.97&
105.73&
113.45&
118.73&
126.65&
133.08&
140.93&
M\"{o}ller \cite{mol95}
\\
94.52&
&
107.04&
&
120.47&
&
&
&
Patyk \cite{pat91}
\\
94.38&
&
106.93&
&
120.47&
&
135.46&
&
Cwiok \cite{cwi94}
\\
\hline
\end{tabular}
\end{center}
\end{scriptsize}
\end{table}

\begin{table}[t]
\begin{center}\begin{small}
\baselineskip=14pt
\centering\caption{Parameters for  even Xe isotopes.
\protect\label{tb:xe1}}

\vspace{0.2em}
\begin{tabular}{cccc}
\heavyline
  Nuclei  &  $ g_6$(keV) &
 $ g'_5$(keV) & $ g'_I$(keV) \\
\heavyline
 $ ^{120}$Xe &  -60   &  53  &  11.9 \\
 $ ^{122}$Xe &  -64   &  59  &  11.9 \\
 $ ^{124}$Xe &  -68.8 &  64  &  11.9 \\
 $ ^{126}$Xe &  -73.3 &  71  &  11.9 \\
 $ ^{128}$Xe &  -78.2 &  79  &  11.9 \\
 $ ^{130}$Xe &  -100.9 & 100  &  11.9 \\
 $ ^{132}$Xe &  -106.5 & 122  &  11.9 \\ \hline
\end{tabular}
\end{small}\end{center}

\begin{center}\begin{small}
\centering\caption{Parameters for odd Xe isotopes.
\protect\label{tb:xe2}}

\vspace{0.2em}
\begin{tabular}{ccccc}
\heavyline
  Nuclei  & $ g_6$(keV) &
 $ g'_5$(keV) & $ g_5$(keV) & $g_J$(keV) \\
\heavyline
 $ ^{127}$Xe & -73.3  &  71.0  & -38.0 & 25.0 \\
 $ ^{129}$Xe & -78.2  &  79.0  & -18.0 & 35.3 \\
 $ ^{131}$Xe & -100.9 & 100.0  &  30.0 & 35.3 \\
 $ ^{133}$Xe & -106.5 & 122.0  &  50.5 & 35.3 \\
 $ ^{135}$Xe &        & 142.1  &  70.6 & 35.3 \\ \hline
\end{tabular}
\end{small}\end{center}

\begin{center}\begin{small}
\centering\caption{Parameters for  odd Ba isotopes.
\protect\label{xe3}}

\vspace{0.2em}
\begin{tabular}{cccc}
\heavyline
  Nuclei  & $ g^i_5$(keV) & $ g_5$(keV) & $g_J$(keV) \\
\heavyline
 $ ^{131}$Ba & 72.5  &  -20  &  35.3 \\
 $ ^{133}$Ba & 105 &  -15  &  35.3 \\
 $ ^{135}$Ba & 90  &   50  &  35.3 \\
 $ ^{137}$Ba & 72  &   70  &  35.3 \\ \hline
\end{tabular}
\end{small}\end{center}
\end{table}

\begin{table}[tbh]

\begin{center}\begin{small}
\centering\caption{Relative $B(E2)$ values for the even Xe isotopes.
\protect\label{xebe2}}

\vspace{0.2em}
\begin{tabular}{rllllllllll}
\heavyline
    & \multicolumn{2}{c}{$^{120}$Xe}
    & \multicolumn{2}{c}{$^{124}$Xe}
    & \multicolumn{2}{c}{$^{126}$Xe}
    & \multicolumn{2}{c}{$^{128}$Xe}
    & \multicolumn{2}{c}{$^{130}$Xe} \\
 $ J_i \rightarrow J_f $ & Exp. & Theo. &  Exp. & Theo.
 &  Exp. & Theo. &  Exp. & Theo. &  Exp. & Theo. \\
\heavyline
 $ 2^+_2 \rightarrow 2^+_1 $ & 100  & 100 & 100 & 100
 & 100 & 100 & 100 & 100 & 100 & 100  \\
 $       \rightarrow 0^+_1 $ & 5.6  & 5.6 & 3.9 & 3.9
 & 1.5 & 1.4 & 1.2 & 1.2 & 0.6 & 0.6  \\
 $ 3^+_1 \rightarrow 2^+_2 $ & 100  & 100 & 100 & 100
 & 100 & 100 & 100 & 100 & 100 & 100  \\
 $       \rightarrow 4^+_1 $ & 50   & 40  & 46  & 40
 & 34  & 40  & 37  & 40  & 25  & 40   \\
 $       \rightarrow 2^+_1 $ & 2.7  & 7.1 & 1.6 & 4.9
 & 2.0 & 1.85 & 1.0 & 1.5 & 1.4 & 0.72  \\
 $ 4^+_2 \rightarrow 2^+_2 $ & 100  & 100 & 100 & 100
 & 100 & 100 & 100 & 100 & 100 & 100  \\
 $       \rightarrow 4^+_1 $ & 62   & 91  & 91  & 91
 & 76  & 91  & 133 & 91  & 107 & 91   \\
 $       \rightarrow 2^+_1 $ & --- & 7.11 & 0.4 & 4.91
 & 1.0 & 1.83 & 1.7 & 1.49 & 3.2 & 0.97  \\
 $ 0^+_2 \rightarrow 2^+_2 $ & 100  & 100 & 100 & 100
 & 100 & 100 & 100 & 100 & 100 & 100  \\
 $       \rightarrow 2^+_1 $ & --- & 7.11 & 1   & 4.91
 & 7.7 & 1.83 & 14  & 1.49 & 26  & 0.97  \\ \hline
\end{tabular}
\end{small}\end{center}
\end{table}

\begin{table}[tbh]
\renewcommand{\arraystretch}{1.5}
\begin{center}\begin{small}
\centering\caption{Transition probablities in $^{129}$Xe ($N_1=5$) and
 $^{131}$Xe ($N_1$=4), $\Omega_1=20$.
\protect\label{xe129-130}}

\vspace{0.2em}
\begin{tabular}{clll|clll}
\heavyline
 \multicolumn{4}{c|}{$^{129}$Xe} &
 \multicolumn{4}{c}{$^{131}$Xe} \\
\heavyline
   & \multicolumn{3}{c|}{$B(E2) (e^2b^2)$} &
   & \multicolumn{3}{c}{$B(E2) (e^2b^2)$} \\ \cline{2-4} \cline{6-8}
 $ J_i \rightarrow J_f $ & FDSM  & ~~Exp. & Ref.\ \reference{jol87} &
 $ J_i \rightarrow J_f $ & FDSM  & ~~Exp. & Ref.\ \reference{jol87} \\
\heavyline
 $ \frac{3}{2}^+_{_{1}} \rightarrow \frac{1}{2}^+_{_{1}} $
 & 0.036 &      & 0.007   &
 $ \frac{1}{2}^+_{_{1}} \rightarrow \frac{3}{2}^+_{_{1}} $
 & 0.0953  & ~~0.0039 & 0.0012 \\

 $ \frac{3}{2}^+_{_{2}} \rightarrow \frac{3}{2}^+_{_{1}} $
 & 0.0186  & $<$0.0005 & 0.013 &
 $ \frac{5}{2}^+_{_{1}} \rightarrow \frac{1}{2}^+_{_{1}} $
 & 0.075  & ~~0.030 & 0.016 \\

 $ \frac{3}{2}^+_{_{2}} \rightarrow \frac{1}{2}^+_{_{1}} $
 & 0.084  & ~~0.12 & 0.12 &
 $ \frac{5}{2}^+_{_{1}} \rightarrow \frac{3}{2}^+_{_{1}} $
 & 0.004  & ~~0.10 & 0.10  \\

 $ \frac{5}{2}^+_{_{1}} \rightarrow \frac{3}{2}^+_{_{1}} $
 & 0.011  & ~~0.22 & 0.10 &
 $ \frac{3}{2}^+_{_{2}} \rightarrow \frac{3}{2}^+_{_{1}} $
 & 0.053  & ~~0.057 & 0.058 \\

 $ \frac{5}{2}^+_{_{1}} \rightarrow \frac{1}{2}^+_{_{1}} $
 & 0.070  & ~~0.077 & 0.039 &
 $ \frac{1}{2}^+_{_{2}} \rightarrow \frac{1}{2}^+_{_{1}} $
 & 0.0000  &          \\

 $ \frac{1}{2}^+_{_{2}} \rightarrow \frac{3}{2}^+_{_{2}} $
 & 0.028  &     &    &
 $ \frac{1}{2}^+_{_{2}} \rightarrow \frac{3}{2}^+_{_{1}} $
 & 0.124  & ~~0.048 & 0.115 \\

 $ \frac{1}{2}^+_{_{2}} \rightarrow \frac{3}{2}^+_{_{1}} $
 & 0.14  & ~~0.044 & 0.12 &
 $ \frac{7}{2}^+_{_{1}} \rightarrow \frac{5}{2}^+_{_{1}} $
 & 0.028  & ~~0.005 & 0.0013 \\

 $ \frac{1}{2}^+_{_{2}} \rightarrow \frac{1}{2}^+_{_{1}} $
 & 0.0000  &   &      &
 $ \frac{7}{2}^+_{_{1}} \rightarrow \frac{3}{2}^+_{_{1}} $
 & 0.043  & ~~0.081 & 0.082  \\

 $ \frac{5}{2}^+_{_{2}} \rightarrow \frac{1}{2}^+_{_{1}} $
 & 0.004  & ~~0.057 & 0.071 &
 $ \frac{3}{2}^+_{_{3}} \rightarrow \frac{3}{2}^+_{_{1}} $
 & 0.025  & ~~0.027 & 0.017 \\

 $ \frac{3}{2}^+_{_{3}} \rightarrow \frac{1}{2}^+_{_{1}} $
 & 0.056  & ~~0.0032 & 0.0056 &
 $ \frac{5}{2}^+_{_{2}} \rightarrow \frac{3}{2}^+_{_{2}} $
 & 0.004  & $<$0.031 & 0.0011 \\

 $ \frac{3}{2}^+_{_{4}} \rightarrow \frac{1}{2}^+_{_{1}} $
 & 0.0133  & ~~0.0030 & 0.0004 &
 $ \frac{5}{2}^+_{_{2}} \rightarrow \frac{1}{2}^+_{_{1}} $
 & 0.071  & ~~0.068 & 0.056  \\

   &   &   &   &
 $ \frac{5}{2}^+_{_{2}} \rightarrow \frac{3}{2}^+_{_{1}} $
 & 0.014  & ~~0.013 & 0.043 \\
   &   &   &   &
 $ \frac{7}{2}^+_{_{2}} \rightarrow \frac{3}{2}^+_{_{1}} $
 & 0.124  & ~~0.005 & 0.026 \\ \hline
\end{tabular}
\end{small}\end{center}
\end{table}

\begin{table}[tbh]
\begin{center}\begin{small}
\centering\caption{Parameters for the Pt calculations.
\protect\label{tb:ptparams}}
\vspace{0.2em}
\begin{tabular}{lrrrrr}
\heavyline
 Mass=   & 190 & 192 & 194 & 196 & 198 \\
\heavyline
$N_{1\nu}$ &  6  &  6  & 5   &   4 &  3 \\  \hline
$B_{2\pi \nu}$ & $-386$  &$ -386$ & $-386$ & $-386$ &$ -386$ \\
$B_{2\pi}$ & 48  & 48  & 48  & 48  & 48 \\
$G_{0\nu}$ & $-49$ & $-49$ & $-49$ & $-49$ & $-49$ \\
$G_{0\pi}$ &$ -10$ & $-3$  & $-5$  & $-18$ & $-50$ \\
$B_{2\nu}$ & 36  & 53  & 74  & 97  & 100 \\ \hline
$e_{\pi}$ & 0.17  & 0.17  & 0.15  & 0.16  & 0.18 \\
$e_{\nu}$ & 0.19  & 0.19  & 0.15  & 0.14  & 0.19 \\ \hline
  \multicolumn{5}{c}{$g_{\pi}=0.62,~~~~~~~~~~~~~~~~g_{\nu}=0$} \\ \hline
  \multicolumn{5}{c}{$\beta_{0\pi}=13\times 10^{-3}\mbox{fm},
  ~~~~\beta_{0\nu}=-1.2\times 10^{-3}\mbox{fm}$}
\\
\hline
\end{tabular}
\end{small}\end{center}
\end{table}

\begin{table}[tbh]
\begin{center}\begin{small}
\centering\caption{Comparison of $B(E2)$ values for $^{196}$Pt.
\protect\label{tb:pt196be2}}
\vspace{0.2em}
\begin{tabular}{rlllllll}
\heavyline
$J_i \rightarrow J_f$
& $B(E2)_{\scriptsize\rm exp}^{(a)}$ & $B(E2)^{(d)}_{\scriptsize\rm exp}$
& $SO(6)$ limit$^{(f)}$ & IBM--2$^{(e)}$
& IBM--1(g) & FDSM \\
\heavyline
$2_1^+ \rightarrow 0_1^+$ &
0.288(14)  & 0.274(1)  &0.288  & 0.289 &0.288 & 0.195 \\
$4_1^+ \rightarrow 2_1^+$ &
0.403(32)  & 0.410(6)  & 0.378 & 0.395 &0.393& 0.248 \\
$6_1^+ \rightarrow 4_1^+$ &
0.421(116) & 0.450(28) &0.384  &0.409 &0.423 & 0.215 \\
$8_1^+ \rightarrow 6_1^+$ &
0.577(58)  &---&0.341  & 0.325 & 0.416 & 0.139 \\
$2_2^+ \rightarrow 2_1^+$ &
0.350(31) & 0.370(5) &0.378 & 0.40 & 0.303 & 0.262  \\
$2_2^+ \rightarrow 0_1^+$ &
$<2.0 \times 10^{-6}$ $^{(b)}$ &--- &0 & 0.001
&0.004& 0.0001 \\
$0_2^+ \rightarrow 2_2^+$ &
0.142(77) &0.1(1) &0.385 & 0.466 & 0.375 & 0.268 \\
$0_2^+ \rightarrow 2_1^+$ &
0.033(7) $^{(b)}$ &0.028(5) &0 &0.026 & 0.007 & 0.021 \\
$4_2^+ \rightarrow 4_1^+$ &
0.193(97) $^{(b)}$ & 0.084(14) &0.183 &0.206 & 0.171 & 0.121 \\
$4_2^+ \rightarrow 2_2^+$ &
0.177(35) $^{(b)}$ &0.18(2) &0.201 &0.206 & 0.199 & 0.112 \\
$4_2^+ \rightarrow 2_1^+$ &
0.0030(10) $^{(b)}$ &0.001(2) &0 &0.006 & 0.004 & 0.001 \\
$6_2^+ \rightarrow 6_1^+$ &
 0.085(121) $^{(b)}$&---&0.108 &0.12& 0.11& 0.075 \\
$6_2^+ \rightarrow 4_2^+$ &
0.350(102) $^{(b)}$ &--&0.232 &0.201 & 0.25 & 0.056 \\
$6_2^+ \rightarrow 4_1^+$ &
0.0037(16) $^{(b)}$ &-&0 &0.017& 0.001& 0.031 \\
$2_{_{3}}^+ \rightarrow 2_1^+$ &
0.0009(15) $^{(b)}$ &--&0
& 0.004&$7.2\times 10^{-6}$& 0.004 \\
$0_{_{3}}^+ \rightarrow 2_1^+$ &
$<0.034$ $^{(c)}$&---&0 & 0.067 & 0.003 &0.010
\\
\hline
\multicolumn{7}{l}{\footnotesize $^{(a)}$
$B(E2)_{\scriptsize\rm exp}$ are taken from H. H. Bolotin {\em et al}.,
Nucl.\ Phys.\ A370 (1981)146}.
\\
\multicolumn{7}{l}{\footnotesize $^{(b)}$
Data are from A. Mauthofer {\em et al}.,  Z. Phys.\
A336 (1990) 263.}
\\
\multicolumn{7}{l}{\footnotesize $^{(c)}$ Data are from
H. G. B\"{o}rner {\em et al}.,
Phys.\ Rev.\ C42 (1990)R2271.}
\\
\multicolumn{7}{l}{\footnotesize $^{(d)}$
$ B(E2)_{\scriptsize\rm exp} $ are taken from C. S. Lim {\em et al}.,
Nucl.\  Phys., A548 (1992) 308.}
\\
\multicolumn{7}{l}{\footnotesize $^{(e)}$
Calc.\ from Bijker {\em et al}., Nucl.\ Phys.\
A344 (1980) 207.}
\\
\multicolumn{7}{l}{$^{(f)}$
For $O(6)$ limit, $e=0.155(eb)$.}
\label{table6}
\end{tabular}
\end{small}\end{center}
\end{table}

\begin{table}[tbh]
\begin{center}
\begin{footnotesize}
\centering\caption {Comparison of experimental and theoretical g-factors in
$^{190-198}$Pt. \protect\label{tb:gfactors}}
\vspace{0.2em}
\begin{tabular}{clllllllll}
\heavyline
& & $2^{+}_{1}$ & & & $4^{+}_{1}$ & &
& $2^{+}_{2}$ &  \\ \cline{2-4} \cline{5-7} \cline{8-10}
isotope & exp.$^{(a)}$ & IBM$^{(c)}$  & FDSM
& exp.$^{(a)}$ & IBM & FDSM
& exp.$^{(a)}$ & IBM & FDSM \\
\heavyline
190 &           & & 0.137 &          & & 0.092 &          & & 0.121 \\
192 & 0.318(17) & 0.23 & 0.182 &0.281(30) & & 0.126 &0.278(46) & & 0.139 \\
194 & 0.295(10) & 0.27 & 0.239 &0.279(31) & & 0.171 &0.281(55) & & 0.178 \\
196 & 0.295(10) & 0.32 & 0.268 &0.345(40) & & 0.213 &0.271(45)$^{(b)}$
   & & 0.207  \\
198 & 0.293(34)$^{(b)}$ & 0.40 & 0.26
      &0.307(54)$^{(b)}$  & & 0.170 &0.307(54)$^{(b)}$
   & & 0.213
\\
\hline
\multicolumn{10}{l}{\footnotesize $^{(a)}$
 Data are taken from
F. Brandolini et al., Nucl.\ Phys.\ A536 (1992) 366.}
\\
\multicolumn{10}{l}{\footnotesize $^{(b)}$
Data are taken from
A. E. Stuchbery, G. J. Lampard and H. H. Bolotin,}
\\[-4pt]
\multicolumn{10}{l}
{\footnotesize \phantom{\footnotesize $^{(b)}$} Nucl.\ Phys.\ A365 (1981) 317;
A528 (1991) 447.}
\\
\multicolumn{10}{l}{\footnotesize $^{(c)}$
IBA--2 calc.\ from M. Sambataro, {\em et al.},
Nucl.\ Phys.\ A423 (1984) 333.}
\label{table8}
\end{tabular}
\end{footnotesize}
\end{center}
\end{table}

\begin{table}[t]
\begin{scriptsize}
\begin{center}
\centering\caption{Experimental and theoretical changes
 of charge radii for $^{190-198}$Pt isotopes. \protect\label{tb:isomer}}
\begin{tabular}{lcccccccccc|cccc}
\heavyline
& \multicolumn{10}{c|}{$\delta <r^2> (10^{-3}\mbox{ fm})$} &
  \multicolumn{4}{c}{$\Delta <r^2> (10^{-3}\mbox{ fm})$} \\
\heavyline
& \multicolumn{2}{c}{190} & \multicolumn{2}{c}{192} &
 \multicolumn{2}{c}{194} & \multicolumn{2}{c}{196} &
 \multicolumn{2}{c|}{198} & 192-190 & 194-192 & 196-194 & 198-196 \\
\hline
& $ 2^+_1 $ & $ 2^+_2 $ & $ 2^+_1 $ & $ 2^+_2 $ &
$ 2^+_1 $ & $ 2^+_2 $ & $ 2^+_1 $ & $ 2^+_2 $ &
$ 2^+_1 $ & $ 2^+_2 $ &           &           & &  \\ \hline
FDSM & 4.48  & 3.22  & 3.84 & 4.76 & 3.16 & 4.93 & 2.66 & 5.54 & 2.14
& 3.68 & 74  & 75 & 71 & 69 \\  
IBM&       &       &      &      & 3.62 & 5.08 & 2.55 &      &
&      &     &    &    &    \\  
Exp.$^{(a)}$  &       &       &      &      &
  3.45 & 3.57 & 4.49 &      &
&      & 65  & 67 & 74 & 80
\\
\hline
\multicolumn{15}{l}{\footnotesize $^{(a)}$
Isomer shifts are taken from R. Engfer {\em et al.},
Nucl.\ Data Tables, 14 (1974) 509;}
\\[-2pt]
\multicolumn{15}{l}{\footnotesize \phantom{$^{(a)}$}
Isotope shifts are taken from
 Th.\ Hilberath {\em et al.}, Z. Phys., A342 (1992) 1.}
\label{table9}
\end{tabular}
\end{center}
\end{scriptsize}
\end{table}

\end{document}